\documentclass[preprint,review,3p,fleqn]{elsarticle}

\usepackage{amssymb, amsthm, amsmath, amsfonts}
\usepackage{graphicx,color}
\usepackage{tabularx}
\usepackage{tikz}
\usepackage[ruled]{algorithm2e}
\usepackage{algorithmic}
\usepackage{subfigure}
\usepackage{lineno}
\usepackage{bm}
\usepackage{multirow,multicol}
\usepackage{epsfig}
\usepackage{epstopdf}
\usepackage{verbatim}
\usepackage{times}
\usepackage{microtype}

\usepackage[colorlinks=true]{hyperref}

\theoremstyle{definition} % amsthm only

\newtheorem{remark}{Remark}

\newcommand{\pa}{\partial}
\newcommand{\f}{\frac}
\newcommand{\nb}{\nabla}

\def\bu{{\bf u}}

\def\a{\alpha}
\def\b{\beta}

\def\XS#1{\textcolor[rgb]{1.00,0.00,0.00}{#1}}
\usepackage{enumitem}
\begin{document}

\begin{frontmatter}

%% Title, authors and addresses

%% use the tnoteref command within \title for footnotes;
%% use the tnotetext command for theassociated footnote;
%% use the fnref command within \author or \address for footnotes;
%% use the fntext command for theassociated footnote;
%% use the corref command within \author for corresponding author footnotes;
%% use the cortext command for theassociated footnote;
%% use the ead command for the email address,
%% and the form \ead[url] for the home page:
%% \title{Title\tnoteref{label1}}
%% \tnotetext[label1]{}
%% \author{Name\corref{cor1}\fnref{label2}}
%% \ead{email address}
%% \ead[url]{home page}
%% \fntext[label2]{}
%% \cortext[cor1]{}
%% \address{Address\fnref{label3}}
%% \fntext[label3]{}

\title{A non-intrusive reduced-order modeling method using polynomial chaos
expansion}

\author[label1]{Xiang Sun}
\author[label1]{Xiaomin Pan}
\author[label1]{Jung-Il Choi\corref{cor1}}\ead{jic@yonsei.ac.kr}

\cortext[cor1]{Corresponding author.}
\address[label1]{Department of Computational Science and Engineering, Yonsei University, Seoul 03722, Republic of Korea}

\begin{abstract}
We propose a non-intrusive reduced-order modeling method based on proper orthogonal decomposition (POD) and polynomial chaos expansion (PCE) for stochastic representations in uncertainty quantification (UQ) analysis. Firstly, POD provides an optimally ordered basis from a set of selected full-order snapshots. Truncating this optimal basis, we construct a reduced-order model with undetermined coefficients. Then, PCE is utilized to approximate the coefficients of the truncated basis. In the proposed method, we construct a PCE using a non-intrusive regression-based method. Combined with the model reduction ability of POD, the proposed method efficiently provides stochastic representations in UQ analysis. To investigate the performance of the proposed method, we provide three numerical examples, i.e., a highly nonlinear analytical function with three uncertain parameters, two-dimensional (2D) heat-driven cavity flow with a stochastic boundary temperature, and 2D heat diffusion with stochastic conductivity. The results demonstrate that the proposed method significantly reduces the computational costs and storage requirements that arise due to high-dimensional physical and random spaces. While demonstrating a similar accuracy with that of the classical full-PCE in predicting statistical quantities. Furthermore, the proposed method reasonably predict the outputs of the full order model using only a few snapshots.
\end{abstract}


\end{frontmatter}

%%%%%%%%%%%%%%%%%%%%%%%%%%%%%%%%%%%%%%%%%
\section{Introduction}
Uncertainty quantification (UQ) is the process of quantifying the effects of input uncertainties on system responses. Recently, UQ has been widely explored in many fields of science and engineering, such as fluid \cite{Najm2009,MK2010,Sun2017} and structural mechanics \cite{Ghanem2007,Schueller2007}. To reproduce the fine-scale structures of such physical phenomena, a large number of degree of freedom (DOF) for simulations is required, which clearly commensurates with increasing computational cost. Hence, for UQ, problems are in general too large to be tackled using standard techniques. This is because approximating system responses with a large number of random parameters on a fine mesh is usually computationally challenging. This emphasizes the need for establishing a model reduction method to reduce computational costs and storage requirements in UQ analysis of the problems with high-dimensional physical and random spaces.

A fundamental problem in UQ is approximating a computational model with random parameters. To this end, many numerical methods have been well developed in recent years (see \cite{Ghanem1991,Helton2003,Xiu2010,Blatman2011,Yadav2014,Schobi2015,Abraham2017} and references therein). One of the most widely used techniques in UQ is polynomial chaos expansion (PCE), which was first proposed by Ghanem and Spanos as finite-dimensional Wiener polynomial chaos \cite{Ghanem1991,Wiener1938}. Later, Xiu and Karniadakis extended \XS{it} to other random variables with basis functions from the Askey family of hypergeometric polynomials \cite{Xiu2002}, which is known as generalized PCE. Furthermore, PCE was extended for random variables with arbitrary distributions in \cite{Witteveen2007,Oladyshkin2012}. To obtain deterministic coefficients of the PCE, two kinds of approaches are used: intrusive and non-intrusive methods. Intrusive methods need solvers for the resulting coupled deterministic equations, which can be very complicated if the form of the equations is nontrivial and nonlinear. However, in the non-intrusive methods, existing solvers are considered as 'black box' without any modifications of the underlying computer codes. The PCE methods always exhibit high convergence rates with increasing expansion order, provided that the solutions are sufficiently smooth with respect to the random variables. However, a well-known shortcoming of these methods is the "curse of dimensionality," i.e., the simulation cost grows rapidly with increasing dimensions of random space. To mitigate this issue, various methods have been developed based on sampling strategies \cite{Xiu2005,Narayan2012,Jakeman2017} and basis reduction \cite{Blatman2011,Abraham2017}. Nevertheless, for problems with high-dimensional physical space, constructing a PCE is still computationally challenging. This is because the computational cost depends not only on the cost of the PCE construction, but also on the number of spatial DOFs. An efficient reduced-order modeling (ROM) method is thus essential for such problems.

ROM \cite{Sirovich1987} aims to approximate the original model using an accurate reduced-order model with a much smaller number of DOFs. Similar to the PCE methods, ROM methods can be classified into two categories in terms of dependency on the governing equations, namely, intrusive and non-intrusive \cite{Frangos2010}. Over the previous decades, many methods have been developed for ROMs (see \cite{Schilders2008,Benner2015,Xiao2016} and references therein). In the present study, we focus on non-intrusive ROM methods based on proper orthogonal decomposition (POD) \cite{Kosambi1943}. Usually, POD-based ROM methods construct a set of basis functions from the snapshots of the original full order model and then approximate the model responses with these basis functions. To this end, a variety of methods have been introduced for POD-based ROM. Audouze et al. \cite{Audouze2009, Audouze2013} proposed a non-intrusive POD-based ROM using radial basis functions (RBFs) for approximating solutions of nonlinear time-dependent parameterized partial differential equations. Xiao et al. \cite{Xiao2010} proposed a non-intrusive ROM based on constrained POD and the Kriging interpolation method, where they used Kriging interpolation to compute the POD coefficients. Guo and Hesthaven \cite{Guo2018,Guo2019} approximated the POD coefficients using Gaussian process regression in ROMs for nonlinear structural analysis and time-dependent problems. Noack et al. \cite{Noack2011} introduced neural networks for ROMs. Later, a non-intrusive ROM method based on POD and RBF artificial neural networks (ANNs) was proposed by Vasile et al. \cite{Vasile2013}. Furthermore, ANNs, particularly multi-layer perceptrons, were introduced for ROMs to accurately approximate coefficients of the reduced model \cite{Hesthaven2018,Wang2018}. Recently, a more extensive and comprehensive discussion on the use of machine learning for the approximation of POD coefficients was proposed in \cite{Swischuk2018}. For more reviews of non-intrusive POD based ROM methods, please refer to \cite{Xiao2016}. In the present study, we focus on developing a non-intrusive ROM method for stochastic representations in UQ problems based on POD and PCE methods. %\XS{The aforementioned methods are capable of being stochastic representations, however, they may not estimate the statistical quantities as easily as PCE methods. We are interested in finding a ROM method that inherits the metrics of PCE methods.}   %

Recently, some algorithms have been proposed based on the POD and PCE methods (see \cite{Doostan2007,Raisee2015,Kumar2016,Abraham2018}). In \cite{Doostan2007}, Doostan et al. proposed an intrusive stochastic model reduction method for creating an alternative optional basis for the PCE by adapting the Karhunen--Lov\`{e}ve (KL) expansion \cite{Loeve1977}. More precisely, a stochastic basis was first obtained on a coarse mesh by means of a KL expansion of the coarse-scale response and then used as the basis functions in fine discretization analysis. According to the results of the numerical examples in \cite{Ghanem2007} and \cite{Doostan2007}, dramatic reduction of computational cost was achieved with little loss in accuracy. However, the method is intrusive, which may be very complicated or even impossible to implement in some cases. To mitigate this problem, Raise et al. \cite{Raisee2015,Kumar2016} extended this idea to a non-intrusive PCE method, which is more efficient and broadly applicable. Moreover, to further reduce the computational cost of PCE construction, Abraham et al. \cite{Abraham2018} further extended this non-intrusive method by introducing a sparse polynomial chaos expansion method \cite{Abraham2017}. Clearly, the new non-intrusive sparse PCE-based model reduction scheme brings a significant speed-up for a wide range of engineering problems, especially for those with high-dimensional random space. However, a full PCE still needs to be constructed on a coarse discretization, which is a large computational challenge for some problems with high resolution in spatial discretization.

In this study, a non-intrusive ROM method is proposed for UQ problems with high-dimensional physical and random spaces. The method combines the advantages of both the POD and PCE methods. However, in comparison with \cite{Gennaro2015}, where POD was first constructed from a data set describing the ice shapes as a reduced-order model and then PCE was constructed based on this model to study the effects of uncertain parameters in POD representation, in our method, the POD basis is directly constructed for fine discretization using the method of snapshots \cite{Sirovich1987}. Then, a sparse PCE method proposed in \cite{Blatman2011} is used to approximate the random coefficients of the POD representation. The method is referred to as POD--PCE. To investigate the performance of this method, three numerical examples were tested, namely, a highly irregular Ackley function with three random parameters, two-dimensional (2D) heat-driven cavity flow with a stochastic boundary temperature, and 2D heat diffusion with stochastic conductivity. %The results indicate the efficiency and accuracy of the POD-PCE in the representation of the stochastic field data computed from full order model with high spatial discretization and high random dimensionality. Moreover, similar to the classical full PCE methods, the statistical quantities can be directly calculated from the deterministic coefficients and POD basis of the POD-PCE representation.

The rest of this paper is organized as follows. In Section 2, we present the POD--PCE method for the representation of the stochastic field, where the construction of the POD basis based on the method of snapshots is discussed, and a sparse PCE method is introduced to approximate the coefficients of the POD representation. Section 3 is dedicated to the numerical tests of the POD--PCE method. Finally, in Section 4 we provide our conclusions.% and future research perspectives.

%%%%%%%%%%%%%%%%%%%%%%%%%%%%%%%%%%%%%%%%%
\section{Representation of the random field}
%Let $(\Omega,\mathcal{F}, P)$ be a probability space with an event space $\Omega$, its $\sigma$-algebra $\mathcal{F}\subset 2^{\Omega}$ and its probability measure $P$. Let $\bm{\xi}=(\xi_1,\xi_2,\ldots,\xi_d)$ be a random vector of $d$ independent random variables for $\omega\in\Omega$.

We consider a random-field model response $U(\bm{x};\bm{\xi})$, where $\bm{\xi}=(\xi_1,\ldots,\xi_d)\in\Gamma=\prod_{k=1}^d\Gamma_k\subset\mathbb{R}^d$ is a random vector with independent components described by the joint probability density function (PDF) $\bm{\rho}(\bm{\xi})=\prod_{k=1}^d \rho_k(\xi_k)$, where $\rho_k(\xi_k)$ is the PDF of the $k$-th component of the random vector. $\bm{x}$ is a spatial vector, which varies in a compact set $\Omega$ of $\mathbb{R}^{s}\ (s=1\ \textrm{or}\ 2\ \textrm{or}\ 3)$. In the present study, we focus on the case of $s=2$. Thus, the variable $\bm{x}$ is deterministic and varies on a grid of size $N$ corresponding to a discretization of $\Omega$. Therefore, $U(\bm{x};\bm{\xi})$ is an $N$-dimensional output given by
\begin{equation}\label{Outputs}
  \bm{U}=\left[U^{(1)},U^{(2)},\ldots,U^{(N)}\right]^T,
\end{equation}
where $U^{(n)}=U(\bm{x}^{(n)};\bm{\xi})$ and $n=1,2,\ldots,N$. Our objective is to develop a reduced-order model for $U(\bm{x};\bm{\xi})$.  The procedure of the construction of the reduced-order model is described in the remainder of this section.

\subsection{Proper orthogonal decomposition}
We first consider a POD method to build a reduced-order model. POD is a powerful and elegant method of data analysis aimed at deriving low-order models of dynamic systems. It was first proposed by Kosambi \cite{Kosambi1943}, and was widely used in various fields of engineering and science under a variety of different names such as Karhunen--Lo\`{e}ve decomposition \cite{Loeve1977} in stochastic process modeling and principal component analysis in statistical analysis \cite{Hotelling1933,Jolliffe2005}. The POD method essentially provides an optimally ordered orthonormal basis in the least-squares sense for a given set of theoretical, experimental, or computational data of high-dimensional systems \cite{Algazi1969}. The optimal lower-dimensional approximations for the given data are then obtained by truncating the optimal basis. In our study, the POD basis vectors are computed empirically using sampled data collected over a subset of the random space $\Omega_r$, typically using the method of snapshots \cite{Sirovich1987}.

Consider an ensemble of snapshots, $\left\{\bm{U}_1, \bm{U}_2, \ldots, \bm{U}_{N_s}\right\}$, where $\bm{U}_{n_s}\in \mathbb{R}^{N}$ denotes the $n_s$-th snapshot, and is given by
\begin{equation}\label{snapshots}
 \bm{U}_{n_s}=\left[U_{n_s}^{(1)}, \ldots, U_{n_s}^{(N)}\right]^T,\ U_{n_s}^{(n)}=U(\bm{x}^{(n)};\bm{\xi}^{(n_s)}),\ n=1,\ldots,N,\ n_s=1,\ldots,N_s.
\end{equation}
Let $\bm{\mathcal{U}}=\left[\bm{U}_1, \bm{U}_2, \ldots, \bm{U}_{N_s}\right]$ be a real-valued snapshot matrix of rank $N_r\leq min\{N,N_s\}$. Then, the POD representation of $U(\bm{x};\bm{\xi})$ is defined as
\begin{equation}\label{POD_Expansion}
  U(\bm{x;\xi})\approx \sum_{l=1}^{L}u_l(\bm{\xi})\psi_l(\bm{x}),
\end{equation}
where $\psi_l(\bm{x})$ are the most characteristic structure functions extracted from the spatial domain $\Omega$ based on the snapshot matrix $\bm{\mathcal{U}}$, $u_l(\bm{\xi})$ are the corresponding coefficients for a given random vector $\bm{\xi}$, and $L\in\{1,\ldots,N_r\}$ is the number of selected structure functions. To determine a proper orthonormal basis $\left\{\bm{\psi}_l\right\}_{l=1}^{L}\ ( \bm{\psi}_l=\left[\psi_l(\bm{x}^{(1)}),\ldots,\psi_l(\bm{x}^{(N)})\right]^T)$ of rank $L$ for the given snapshots, we need to solve the following eigenvalue problem
%Since POD is usually reduced to the following eigenvalue problem
\begin{equation}\label{EignP}
  \bm{\mathcal{U}\mathcal{U}}^T\bm{\psi}=\lambda\bm{\psi},
\end{equation}
which is strongly related to the singular value decomposition (SVD) \cite{Noble1969} of matrix $\bm{\bm{\mathcal{U}}}$; i.e.,
\begin{equation}\label{SVD}
  \bm{\mathcal{U}}=\bm{\Psi\Sigma V}^T, \qquad \bm{\Sigma}=   \begin{pmatrix}
   \bm{D} & \bm{0} \\
   \bm{0} & \bm{0}  \\
  \end{pmatrix},
\end{equation}
where $\bm{D}=diagonal(\sigma_1,\ldots,\sigma_{N_r})\in \mathbb{R}^{N_r\times N_r}$ containing the singular values $\sigma_1\geq \sigma_2\geq \cdots\geq \sigma_{N_r}\geq 0$ and zeros denote matrices of appropriate dimensions with zero-elements. $\bm{\Psi}\in\mathbb{R}^{N\times N}$ and $\bm{V}\in\mathbb{R}^{N_s\times N_s}$ are unitary matrices, i.e., $\bm{\Psi}\bm{\Psi}^T=\bm{\Psi}^T\bm{\Psi}=\bm{I}_{N}$ and $\bm{VV}^T=\bm{V}^T\bm{V}=\bm{I}_{N_s}$.
Let $\bm{\psi}_{n_r}\in\mathbb{R}^{N}$ and $\bm{\nu}_{n_r}\in\mathbb{R}^{N_s}$ be the $n_r$-th column vectors of $\bm{\Psi}$ and $\bm{V}$, respectively. Then we have
\begin{equation}\label{SVD_p}
  \bm{\mathcal{U}}\bm{\nu}_{n_r}=\sigma_{n_r}\bm{\psi}_{n_r}\ \ \textrm{and}\ \  \bm{\mathcal{U}}^T\bm{\psi}_{n_r}=\sigma_{n_r}\bm{\nu}_{n_r}\ \ \textrm{for}\ \ {n_r}=1,\ldots,N_r.
\end{equation}
Thus,
\begin{equation}\label{SVD_pp}
  \bm{\mathcal{U}}^T\bm{\mathcal{U}}\bm{\nu}_{n_r}=\sigma_{n_r}^2\bm{\nu}_{n_r}\ \ \textrm{and}\ \  \bm{\mathcal{U}}\bm{\mathcal{U}}^T\bm{\psi}_{n_r}=\sigma_{n_r}^2\bm{\psi}_{n_r}\ \ \textrm{for}\ \ {n_r}=1,\ldots,N_r.
\end{equation}
Hence, $\left\{\bm{\psi}_{n_r}\right\}_{{n_r}=1}^{N_r}$ and $\left\{\bm{\nu}_{n_r}\right\}_{n_r=1}^{N_r}$ are eigenvectors of $\bm{\mathcal{U}}\bm{\mathcal{U}}^T$ and $\bm{\mathcal{U}}^T \bm{\mathcal{U}}$, respectively, with eigenvalues $\lambda_{n_r}=\sigma_{n_r}^2>0$, ${n_r}=1,\ldots,N_r$.
Under the assumption of $N_s\ll N$, we reformulate the computation of POD modes as an $N_s\times N_s$ eigenvalue problem that is computationally efficient; i.e.,
\begin{equation}\label{EigenPP}
  \bm{\mathcal{U}}^T\bm{\mathcal{U}}\bm{\nu}=\lambda\bm{\nu}.
\end{equation}

According to Eq. (\ref{SVD_p}), $\left\{\bm{\psi}_l\right\}_{l=1}^{L}\ (L\in\{1,\ldots,N_r\},\  \bm{\psi}_l=\bm{\mathcal{U}}\bm{\nu}_l/\sqrt{\lambda_l})$ is the POD basis of rank $L$ that we need \cite{Volkwein2011}. Then, a reduced-order model is obtained by projecting the full order model onto the space spanned by the POD basis of rank $L$. The POD basis is optimal in the sense that it minimizes the least-squares error of the snapshot representation for the reduced basis of size $L$. Let $\widetilde{\bm{\Psi}}\in \mathbb{R}^{N\times L}$ be a matrix with a pairwise orthonormal vector $\bm{\psi}_{l}$. Then, the projection error of the snapshots can be evaluated as
\begin{equation}\label{POD_LS}
\begin{aligned}
  \|\bm{\mathcal{U}}-\widetilde{\bm{\Psi}}\widetilde{\bm{\Psi}}^T\bm{\mathcal{U}}\|_F^2&=\min_{\bm{\Psi}^L\in\mathbb{R}^{N\times L}}\|\bm{\mathcal{U}}-\bm{\Psi}^L(\bm{\Psi}^L)^T\bm{\mathcal{U}}\|_F^2\\
       &=\min_{\bm{\Psi}^L\in\mathbb{R}^{N\times L}}\sum_{n_s=1}^{N_s}\|\bm{U}_{n_s}-\bm{\Psi}^L(\bm{\Psi}^L)^T\bm{U}_{n_s}\|_2^2=
       \sum_{n_s=L+1}^{N_s}\lambda_{n_s}=\sum_{n_s=L+1}^{N_r}\lambda_{n_s},
\end{aligned}
\end{equation}
where $\bm{\Psi}^L=\left[\bm{\psi}_{l_1},\ldots,\bm{\psi}_{l_L}\right]\ (l_i,\in \{1,\ldots,N_r\},i=1,\ldots,L)$, $\|\cdot\|_2$ is the general $L_2$ norm for a vector (e.g.,$\|\bm{\psi}_l\|_2=\sqrt{\sum_{n=1}^{N}\psi_l^2(\bm{x}^{(n)})}$), and $\|\cdot\|_F$ denotes the Frobenius norm given by
\begin{equation}\label{F_norm}
  \|\bm{A}\|_F=\sqrt{\sum_{n_1=1}^{N_1}\sum_{n_2=1}^{N_2}|A_{n_1,n_2}|^2}\qquad \textrm{for} \ \bm{A}\in \mathbb{R}^{N_1\times N_2}.
\end{equation} %=\sqrt{trace(A^TA)}
%For the detail of the equation deducing, please refer to \cite{Volkwein2011}.
%The equation indicates that the POD basis is optimally selected in the least-square sense.
As seen in Eq.~(\ref{POD_LS}), the square of the error in the snapshot representation is defined by the sum of the eigenvalues of the eigenvalue problem (\ref{EigenPP}) corresponding to the modes not included in the selected basis. Furthermore, a fast decay of the eigenvalues of $\bm{\mathcal{U}}^T\bm{\mathcal{U}}$ is usually assumed in POD. Therefore, the size of the POD basis can be chosen based on the size of the dominant eigenspace such that
\begin{equation}\label{Tol}
  \frac{\sum_{l=1}^L\lambda_l}{\sum_{l=1}^{N_r}\lambda_l}>1-\varepsilon,
\end{equation}
where $\varepsilon$ is a user-specified tolerance, often considered to be $10^{-3}$ or smaller \cite{Benner2015}. The numerator of Eq.~(\ref{Tol}) is often referred to as the "energy" captured by the POD modes.

Let
\begin{equation}\nonumber
  V_{N}=\left\{U(\bm{x;\xi}): \bm{x}\in\Omega, \bm{\xi}\in\Gamma\right\},\ V_{N_s}= span\left\{\bm{U}_1, \bm{U}_2, \ldots, \bm{U}_{N_s}\right\},\ \Xi= span\left\{\bm{\xi}^{(1)}, \bm{\xi}^{(2)}, \ldots, \bm{\xi}^{(N_s)}\right\},
\end{equation}
where $V_{N}$ is the realization domain of $U(\bm{x;\xi})$, $V_{N_s}\subset V_{N}$ is spanned by the snapshots and $\Xi$ is a subset of $\Gamma$ spanned by a set of selected random vectors. Then, if the random vectors are selected sufficiently well, $U(\bm{x;\xi})$ can be approximated using the basis of $V_{N_s}$. As mentioned above, the main result of POD is that the optimal subspace $V_L$ of dimension $L\ll \min\{N,N_s\}$ representing the snapshots is given by $V_L=span\left\{\bm{\psi}_{1}, \bm{\psi}_{2}, \ldots, \bm{\psi}_{L}\right\}$. Thus, for any random vector $\bm{\xi}\in \Gamma$, $U(\bm{x;\xi})$ can be approximated as shown in Eq. (\ref{POD_Expansion}).
%\begin{equation}\label{POD_Expansion}
%  U(\bm{x;\xi})\approx\sum_{i=1}^{L}u_i(\bm{\xi})\psi^{(i)}(\bm{x}).
%\end{equation}
Clearly, the coefficients $u_l(\bm{\xi})$ in Eq.~(\ref{POD_Expansion}) vary with the random vector $\bm{\xi}$, which means that we can construct a mapping between the POD coefficients and the random vector $\bm{\xi}$. To this end, a sparse PCE method is introduced in the following subsection.

\begin{remark}\label{remark1}
The POD bases optimally approximate ensemble snapshots of field $U(\bm{x;\xi})$ in the least-squares sense, but they are not bases approximating the field $U(\bm{x;\xi})$. Only when the random vectors are selected well enough can the POD bases be used to approximate the field data. Hence, choosing good random vectors for the snapshots is important in the method of snapshots. In the present study, the snapshots are chosen by a sequential sampling scheme, as described in Section 2.4.
\end{remark}
%%%%%%%%%%%%%%%%%%%%%%%%%%%%%%%%%%%%%
\subsection{Polynomial chaos expansion}
We now present the basic formulation of the stochastic representations of the coefficients $u_l (l=1,\ldots,L)$ in the POD representation of the random field $U(\bm{x};\bm{\xi})$ using the PCE method.

The basic idea behind PCE is to use orthogonal polynomials in terms of random variables to approximate functions of those random variables. As shown in \cite{Xiu2010,Xiu2002}, the polynomial basis $\bm{\phi}^{(k)}(\xi_k) (k=1,\ldots,d)$ can be chosen on the basis of the correspondence between the weighting function of the polynomial family and the PDF of the random variable $\xi_k$; for example, normal and uniform distributions correspond to Hermite and Legendre polynomial bases, respectively. Furthermore, the univariate polynomial bases $\bm{\phi}^{(k)}(\xi_k)$ satisfy the following orthogonality:
\begin{equation}\label{ortho}
  \mathbb{E}\left[\phi_i^{(k)}(\xi_k)\phi_j^{(k)}(\xi_k)\right]=\int_{\Gamma_{k}}\phi_i^{(k)}(\xi_k)\phi_j^{(k)}(\xi_k)\rho_k(\xi_k)d\xi_k=\delta_{ij},
\end{equation}
where $\delta_{ij}$ is the Kronecker delta function. Then, the PCE of $u_l(\bm{\xi})$ is defined as
\begin{equation}\label{PCE}
  u_l(\bm{\xi})=\sum_{\bm{\alpha}\in \mathbb{N}^d}c_{\bm{\alpha}}^{(l)}\Phi_{\bm{\alpha}}(\bm{\xi}),
\end{equation}
where $\Phi_{\bm{\alpha}}(\bm{\xi})=\phi_{\alpha_1}^{(1)}(\xi_1)\cdots \phi_{\alpha_d}^{(d)}(\xi_d)$ are multivariate polynomials orthonormal with respect to $\bm{\rho}(\bm{\xi})$ and $\bm{\alpha}\in \mathbb{N}^d$ is a multi-index that identifies the components of the multivariate polynomials $\Phi_{\bm{\alpha}}$ and  satisfies that $|\bm{\alpha}|=\alpha_1+\cdots+\alpha_d,\ |\bm{\alpha}|\geq 0$. In addition, $c_{\bm{\alpha}}^{(l)}$ are the deterministic coefficients corresponding to the polynomials. The convergence of the Eq.~(\ref{PCE}) should be justified for different polynomials in practice. In short, the Wiener--Hermite chaos \cite{Ghanem1991} can be justified by the Cameron--Martin theorem \cite{CaMa47}. Rigorous justifications about the convergence of generalized PCE can be found in \cite{Ernst12}.

Practically, the sum in Eq.~(\ref{PCE}) needs to be truncated to a finite sum by introducing the truncated polynomial chaos expansion:
\begin{equation}\label{TPCE}
  u_l(\bm{\xi})\approx\sum_{\bm{\alpha}\in \mathcal{A}}c_{\bm{\alpha}}^{(l)}\Phi_{\bm{\alpha}}(\bm{\xi}),
\end{equation}
where $\mathcal{A}\subset \mathbb{N}^d$ is the set of selected multi-indices of multivariate polynomials which may have the following form:
\begin{equation}\label{Truncation}
  \mathcal{A}=\left\{\bm{\alpha}\in \mathbb{N}^d : |\bm{\alpha}|\leq p\right\},\qquad card\{\mathcal{A}\}=M=\frac{(p+d)!}{p!d!}.
\end{equation}
Because of the orthogonality relations in Eq.~(\ref{ortho}), it follows that the multivariate polynomials are also orthonormal:
\begin{equation}\label{ortho_multi}
  \mathbb{E}\left[\Phi_{\bm{\alpha}}(\bm{\xi})\Phi_{\bm{\beta}}(\bm{\xi})\right]=\delta_{\bm{\alpha}\bm{\beta}},
\end{equation}
where $\delta_{\alpha\beta}$ is an extension Kronecker symbol to the multi-dimensional case. Based the orthogonality relations, we can easily estimate the statistical quantities of $u_l$; e.g., for a PCE with the first multivariate polynomial $\Phi_{\bm{0}}(\bm{\xi})=1,$ we have
\begin{equation}\label{mean}
  \mu(u_l)=\mathbb{E}\left[u_l(\bm{\xi})\right]\approx\int_{\Gamma}\left(\sum_{\bm{\alpha}\in \mathcal{A}}c_{\bm{\alpha}}^{(l)}\Phi_{\bm{\alpha}}(\bm{\xi})\right)\bm{\rho}(\bm{\xi})d\bm{\xi}=c_{\bm{0}}^{(l)},
\end{equation}
and
\begin{equation}\label{variance}
  var(u_l)=\mathbb{E}\left[(u_l(\bm{\xi})-\mu(u_l))^2\right]\approx \sum_{0<|\bm{\alpha}|\leq p}(c_{\bm{\alpha}}^{(l)})^2.
\end{equation}

%\begin{equation}\label{OPCE_Cov}
%  C(x_i,x_j)=\sum_{k=1}^{P-1}c_{k}(x_i)c_{k}(x_j).
%\end{equation}

To calculate the coefficients of the PCE for a given polynomial basis, many methods have been proposed, e.g., see \cite{Xiu2010,Blatman2011,Hosder2006,Hosder2007,Doostan2011}. In the present study, only regression-based non-intrusive methods are considered, i.e., the PCE coefficients are estimated by minimizing the mean squared error of the response approximation. A direct approach to estimating the coefficients is to set up a least-squares minimization problem \cite{Berveiller2006}. Thus, Eq.~(\ref{PCE}) can be written as a sum of its truncated version and a residual:
\begin{equation}\label{TPCE_error}
  u_l(\bm{\xi})=\sum_{m=0}^{M-1}c_{k}^{(l)}\Phi_{m}(\bm{\xi})+\varepsilon_M=\bm{\Phi}\bm{c}^{(l)}+\varepsilon_M,
\end{equation}
where $\varepsilon_M$ is the truncation error, $\bm{\Phi}=\left[\Phi_0(\bm{\xi}),\ldots,\Phi_{M-1}(\bm{\xi})\right]$ is the matrix that assembles the values of all the orthonormal polynomials in $\bm{\xi}$, and $\bm{c}^{(l)}=\left[c_0^{(l)},\ldots,c_{M-1}^{(l)}\right]^T$ is a vector containing the coefficients.

The least-squares minimization problem is then set up as:
\begin{equation}\label{OLS}
  \hat{\bm{c}}^{(l)}=\arg\min_{\bm{c}^{(l)}} \mathbb{E}\left[\left(\bm{\Phi}\bm{c}^{(l)}-u_l(\bm{\xi}) \right)^2\right].
\end{equation}
Given a sampling of size $N_{\xi}$ of the random vector $\{\bm{\xi}^{(n_{\xi})}\}_{n_{\xi}=1}^{N_{\xi}}$, we have $\bm{\mathcal{U}}=\left[\bm{U}_1, \bm{U}_2, \ldots, \bm{U}_{N_{\xi}}\right]$, where $N_s=N_{\xi}$. Based on Eq. (\ref{POD_Expansion}), the corresponding coefficients can be estimated as $\bm{u}_l=\bm{\widetilde{\Psi}}^T\bm{\mathcal{U}}=\left[u_l(\bm{\xi}^{(1)}),\ldots,u_l(\bm{\xi}^{(N_{\xi})})\right]^T,$ then
we have
\begin{align}\nonumber
\mathbb{E}\left[\left(\bm{\Phi}\bm{c}^{(l)}-u_l(\bm{\xi}) \right)^2\right]&=\int_{\Gamma}\left(\bm{\Phi}(\bm{\xi})\bm{c}^{(l)}-u_l(\bm{\xi})\right)^2\bm{\rho}(\bm{\xi})d\bm{\xi}\\ \nonumber
&\approx\frac{1}{N_{\xi}}\sum_{n_{\xi}=1}^{N_{\xi}}\left(\bm{\Phi}(\bm{\xi}^{(n_{\xi})})\bm{c}^{(l)}-u_l(\bm{\xi}^{(n_{\xi})})\right)^2
=\frac{1}{N_{\xi}}\|\bm{A}\bm{c}^{(l)}-\bm{u}_l\|_2^2.
\end{align}
where
\begin{equation}\label{LS_Matrix}
  A_{n_{\xi},m}=\Phi_m(\bm{\xi}^{(n_{\xi})}),\ n_{\xi}=1,\ldots,N_{\xi},\ \textrm{and}\ m=0,\ldots,M-1.
\end{equation}
Then, the ordinary least-square solution of Eq.~(\ref{OLS}) is calculated as:
\begin{equation}\label{OLS_solution}
  \hat{\bm{c}}^{(l)}=(\bm{A}^T\bm{A})^{-1}\bm{A}^T\bm{u}_l.
\end{equation}
%where
%\begin{equation}\label{LS_Matrix}
%  A_{jk}=\Phi_k(\bm{\xi}^{(j)}),\qquad j=1,\ldots,N,\ \textrm{and}\ k=0,\ldots,P-1.
%\end{equation}

According to Hosder et al. \cite{Hosder2007}, $2M\sim 3M$ samples are sufficient for accurate approximation to the statistics at each polynomial degree. However, considering this design of experiment (DoE), the computational effort may become untenable for large values of $d$ or $p$; e.g., $d\geq 10$ or $p\geq 10$. To mitigate this issue, several methods have been developed (see \cite{Blatman2011,Abraham2017,Baptista2019} and references therein). In our study, the least-angle regression-based adaptive sparse PCE is adopted, which is developed based on the sparsity of effects; i.e., most computational models are driven by main effects and low-order interactions \cite{Blatman2011}. In fact, the sparse PCE solves a modified least-squares problem in Eq.~(\ref{LS_l1}) by adding a penalty term:
\begin{equation}\label{LS_l1}
  \hat{\bm{c}}^{(l)}=\arg\min_{\bm{c}^{(l)}} \mathbb{E}\left[\left(\bm{\Phi}\bm{c}^{(l)}-u_l(\bm{\xi}) \right)^2\right]+\gamma\|\bm{c}^{(l)}\|_1,
\end{equation}
where $\gamma$ is the regularization parameter and $\|\cdot\|_1$ is the $l_1$ norm that sums the absolute element of a vector.
\begin{remark}\label{remark2}
Conventional PCE methods focus on the problem with smooth model response with regard to input variables. Thus when there are steep gradients or finite discontinuities in the random space, these methods converge very slowly or even fail to converge due to a linear combination of polynomials. For these cases, local expansions may be more efficient than expansions with global polynomials, such as the wavelet expansions \cite{Maitre2004} and the multi-element generalized PCE \cite{Wan2005,Wan2006}. In the present study, we mainly focus on the problems with smooth responses with respect to input variables.
\end{remark}
%Similar to \cite{Blatman2011}, the leave-one-out cross validation error ($\varepsilon_{LOO}$) is accepted to evaluate the predictive ability of a PCE metamodel in this work, which has been proved in \cite{Blatman2011} that it performs well in the accuracy estimation of PCEs. The error can be defined as follows:
%%%
%\begin{equation}\label{LOO}
%  \varepsilon_{LOO}=\frac{\sum_{j=1}^{N}\left(u_i(\bm{\xi}^{(j)})-u_i^{PCE\setminus j}(\bm{\xi}^{(j)})\right)^2}{\sum_{j=1}^{N}\left(u_i(\bm{\xi}^{(j)})-\mu_{u_i}\right)^2},\qquad \mu_{u_i}=\frac{1}{N}\sum_{j=1}^{N}u_i(\bm{\xi}^{(j)})
%\end{equation}
%%%
%Where $u_i^{PCE\setminus j}$ is the corresponding PCE metamodel constructed based on $N$ samples of input uncertain variables excluding the $j$-th sample $\bm{\xi}^{(j)}$.
%%%%%%%%%%%%%%%%%%%%%%%%%%%%%%%%%%%%%%%%%
\subsection{POD--PCE}
Combining the POD representation in Eq.~(\ref{POD_Expansion}) and the PCE approximations in Eq.~(\ref{TPCE}) of the undetermined coefficients of the POD basis, a whole representation of the random field $U(\bm{x};\bm{\xi})$ can be obtained as follows:
%\begin{equation}\label{POD_PCE}
%  U(\bm{x};\bm{\xi})=\sum_{i=1}^{r}\sum_{\bm{\alpha}\in \mathcal{A}}c_{\bm{\alpha}}^{(i)}\Phi_{\bm{\alpha}}(\bm{\xi})\psi^{(i)}(\bm{x}).
%\end{equation}
%More precisely, we have
\begin{equation}\label{POD_PCE_1}
  U(\bm{x};\bm{\xi})\approx\widetilde{U}(\bm{x;\xi})=\sum_{l=1}^{L}\left(\sum_{m=0}^{M-1}c_{m}^{(l)}\Phi_{m}(\bm{\xi})\right)\psi_l(\bm{x}).
\end{equation}
This stochastic representation is referred to as POD--PCE in the present study. The method reduces the spatial dimensionality from $N$ to $L$. Furthermore, similar to the classical PCE methods, it can estimate the statistical quantities of $U(\bm{x};\bm{\xi})$ in terms of $\bm{\xi}$ using the orthogonality of the PCE; more precisely, the statistical quantities can be calculated by PCE coefficients and POD basis functions; e.g.,
 \begin{equation}\label{mean_RF}
  \mu(U(\bm{x};\bm{\xi}))=\mathbb{E}\left[U(\bm{x};\bm{\xi})\right]\approx\mathbb{E}\left[\sum_{l=1}^{L}u_l(\bm{\xi})\psi_{l}(\bm{x})\right]
  =\sum_{l=1}^{L}\mathbb{E}\left[u_l(\bm{\xi})\right]\psi_{l}(\bm{x})=\sum_{l=1}^{L}c_{0}^{(l)}\psi_{l}(\bm{x}),
\end{equation}
and

\begin{align}
 var(U(\bm{x};\bm{\xi}))&\approx var\left[\sum_{l=1}^{L}u_l(\bm{\xi})\psi_{l}(\bm{x})\right]=
 \sum_{l_1=1}^L\sum_{l_2=1}^L\psi_{l_1}(\bm{x})\psi_{l_2}(\bm{x})Cov(u_{l_1},u_{l_2}),\label{variance_RF1}
\end{align}
where
\begin{align}\label{Cov_PCE}
 Cov(u_{l_1},u_{l_2})&=\mathbb{E}\left[(u_{l_1}(\bm{\xi})-\mu(u_{l_1}))(u_{l_2}(\bm{\xi})-\mu(u_{l_2}))\right]\\
       &\approx \mathbb{E}\left[(u_{l_1}(\bm{\xi})-c_0^{(l_1)})(u_{l_2}(\bm{\xi})-c_0^{(l_2)})\right]\\
       &=\sum_{m=1}^{M-1}c_m^{(l_1)}c_m^{(l_2)}.
\end{align}
%and $var(u_l)$ is calculated as shown in Eq.~(\ref{variance}).
\subsection{Design of experiment in POD--PCE}
As presented in the above subsections, a good design of experiment (DoE) is necessary in both the POD and PCE parts. In the POD part, we have to generate a proper set of snapshots such that the space expanded by the POD basis extracted from these snapshots can approximate the random field response. In the PCE part, a proper number of random samples should be generated to obtain a convergent result under some accuracy requirements.

According to the discussion in \cite{Volkwein2011}, the continuous form of the eigenvalue problem (\ref{EigenPP}) can be written as:
\begin{equation}\label{conti_version_1}
  \int_{\Gamma}\left\langle \bm{U},\bm{\psi}\right\rangle_{\mathbb{R}^{N}}U(\bm{x;\xi})\bm{\rho}(\bm{\xi})d\bm{\xi}=\lambda\psi(\bm{x}).
\end{equation}
where $\langle\cdot,\cdot\rangle_{\mathbb{R}^{N}}$ is the inner product in ${\mathbb{R}^{N}}$ and $\left\langle \bm{U},\bm{\psi}\right\rangle_{\mathbb{R}^{N}}=\bm{U}^T\bm{\psi}$.
Then, we have
\begin{equation}\label{conti_version_2}
  \sum_{n_1=1}^{N}\left(\int_{\Gamma} U(\bm{x}^{(n_1)};\bm{\xi}) U(\bm{x}^{(n_2)};\bm{\xi})\bm{\rho}(\bm{\xi})d\bm{\xi}\right) \psi(\bm{x}^{(n_1)})=\lambda\psi(\bm{x}^{(n_2)}),\ n_2=1,\ldots,N.
\end{equation}
Let
\begin{equation}\label{conti_version_Cov}
  C_{n_1,n_2}=\int_{\Gamma} U(\bm{x}^{(n_1)};\bm{\xi}) U(\bm{x}^{(n_2)};\bm{\xi})\bm{\rho}(\bm{\xi})d\bm{\xi}=\mathbb{E}\left[U(\bm{x}^{(n_1)};\bm{\xi}) U(\bm{x}^{(n_2)};\bm{\xi})\right],\ n_1=1,\ldots,N,\ n_2=1,\ldots,N.
\end{equation}
We obtain the following eigenvalue problem:
\begin{equation}\label{conti_version_EP}
  \bm{C}\bm{\psi}=\lambda\bm{\psi},
\end{equation}
where $\bm{C}=\left(C_{n_1,n_2}\right)_{N\times N}$ is the correlation matrix of the random field $U(\bm{x};\bm{\xi})$. In comparison with the eigenvalue problem (\ref{EignP}), $\bm{\mathcal{U} \mathcal{U}}^T$ can be considered as an approximation to $\bm{C}$. The approximation accuracy depends on the method used to solve Eq.~(\ref{conti_version_Cov}), such as the Latin hypercube sampling (LHS) method \cite{Helton2003}, sparse grid \cite{Nobile2008}, and greedy sampling method \cite{Chen2018}. In the present study, we approximate $\bm{C}$ by using the snapshots chosen by a sequential sampling scheme as follows:
\begin{enumerate}
  \item Generate an initial set of random vectors as $\bm{\Theta}_0=\left\{\bm{\xi}^{(1)},\ldots,\bm{\xi}^{(N_0)}\right\}$ using the LHS method \cite{Helton2003}.
  \item Calculate snapshots $\bm{U}_{n_0}$ ($n_0=1,\ldots,N_0$) and assemble the snapshot matrix $\bm{\mathcal{U}}=\left[\bm{U}_1, \ldots, \bm{U}_{N_0}\right]$.
  \item Solve the eigenvalue problem of $\bm{\mathcal{U}^T\mathcal{U}}$ to generate $\lambda_{n_0}^{0}$ and normalize the eigenvalues as $\hat{\lambda}_{n_0}^{0}=\frac{\lambda_{n_0}^{0}}{\lambda_{1}^{0}},\ \lambda_{1}^{0}\geq\cdots\geq \lambda_{N_0}^{0}$.
  \item Enrich $\bm{\Theta}_0$ to be $\bm{\Theta}_1\supset \bm{\Theta}_0$ of size $N_1$ using a nested LHS method \cite{Qian2009} and repeat (2)--(3) to obtain normalized eigenvalues $\left\{\hat{\lambda}_{n_1}^{1}\right\}_{n_1=1}^{N_1}$.
  \item Select $N_l$ eigenvalues such that $\hat{\lambda}_{N_l}^{k}>\tau,\ k=0,1$ and compute the mean absolute relative error as $\varepsilon_{\lambda}=\frac{1}{N_l}\sum_{n_l=1}^{N_l}\left|\frac{\hat{\lambda}_{n_l}^{0}-\hat{\lambda}_{n_l}^{1}}{\hat{\lambda}_{n_l}^{1}}\right|$.
  \item Update $\bm{\Theta}_0=\bm{\Theta}_1,\ N_0=N_1$ and repeat (1)--(5) until $\varepsilon_{\lambda}<\hat{\tau}$. Then define the number of the snapshots as $N_s=N_0$ and select $\mathcal{D}_{POD}=\left\{\left(\bm{\xi}^{(n_s)}, \bm{U}_{n_s}\right),\ n_s=1,2,\dots,N_s\right\}$ as the DoE of the POD construction.
\end{enumerate}
In the above procedure, tolerance $\tau$ is used to select the eigenvalues with large "energy." This is because the decay of the eigenvalue is usually rapid and the smallest eigenvalues have little influence on the total energy. The tolerance can be used to reduce the computation complexity of error $\varepsilon_{\lambda}$. For the tolerance $\hat{\tau}$, it is the stopping criterion of the procedure that ensures that the difference between the eigenvalues computed by the neighbor iterations is small enough. In the present study, we set $\tau=10^{-10}$ and $\hat{\tau}=0.05$ throughout the paper.

For the PCE part, the DoE is defined as $\mathcal{D}_{PCE}=\left\{\left(\bm{\xi}^{(n_s)}, \bm{\widetilde{\Psi}}^T\bm{\mathcal{U}}(\bm{\xi}^{(n_s)})\right),\ n_s=1,2,\dots,N_s\right\}$ (see Section 2.2). This implies that we use the same size DoE for both POD and PCE. This is  because we apply sparse PCE to approximate the coefficients of POD, which can obtain an accurate approximation by only a few samples of input variables. Additionally, the above snapshot selection procedure demonstrates that the final selected snapshots lead to a set of convergent normalized eigenvalues. This indicates that the eigenvalue has the same decay when there are $N_s$ or more snapshots. Thus, the $N_s$ snapshots can capture most of the information in the random field $U(\bm{x};\bm{\xi})$. Therefore, using the same DoE in PCE is reasonable. If the PCE approximation does not satisfy the accuracy requirement under the same DoE as that of POD, we can enrich the samples by using the nested LHS.

%%%%%%%%%%%%%%%%%%%%%%%%%%%%%%%%%%%%%%%%%%%%%%%%%%%%%%%%%%%%%%%%%
\subsection{Error analysis of POD--PCE}
For any given parameters $\bm{\xi}\in \Gamma$, we have
\begin{equation}\label{POD-PCEA}
\begin{aligned}
U(\bm{x;\xi})&=\sum_{l=1}^{L}\left(\sum_{m=0}^{M-1}c_{m}^{(l)}\Phi_{m}(\bm{\xi})+\varepsilon_M^{(l)}\right)\psi_l(\bm{x})+\varepsilon_{L}(\bm{x})\\
             &=\sum_{l=1}^{L}\sum_{m=0}^{M-1}c_{m}^{(l)}\Phi_{m}(\bm{\xi})\psi_l(\bm{x})+\sum_{l=1}^{L}\varepsilon_M^{(l)}\psi_l(\bm{x})+\varepsilon_{L}(\bm{x}).
\end{aligned}
\end{equation}
Here, $\varepsilon_{L}(\bm{x})$ is the error caused by POD truncation and snapshot selection. Let $\bm{\varepsilon}_M=\left[\varepsilon_M^{(1)},\ldots,\varepsilon_M^{(L)}\right]^T$ and $\bm{\varepsilon}_L=\left[\varepsilon_L(\bm{x}^{(1)}),\ldots,\varepsilon_L(\bm{x}^{(N)})\right]^T$.
Then, the error of POD--PCE approximation to $U(\bm{x;\xi})$ for a spatial discretization of $N$ grids is defined as follows: %{\mathbb{R}^{N}}
\begin{equation}\label{PPER}
\begin{aligned}
 \varepsilon_{PP}&=\left\|\bm{U}-\widetilde{\bm{U}}\right\|_2=\left\|\widetilde{\bm{\Psi}}\bm{\varepsilon}_M+\bm{\varepsilon}_{L}\right\|_2\\
  &\leq \left\|\widetilde{\bm{\Psi}}\bm{\varepsilon}_M\right\|_2+\left\| \bm{\varepsilon}_{L} \right\|_2\\
  &\leq \left\|\widetilde{\bm{\Psi}}\right\|_{2,2}\left\|\bm{\varepsilon}_M\right\|_2+\left\| \left(\bm{U}-\widehat{\bm{\Psi}}\widehat{\bm{\Psi}}^T\bm{U}\right)+ \left(\widehat{\bm{\Psi}}\widehat{\bm{\Psi}}^T\bm{U}-\widetilde{\bm{\Psi}}\widetilde{\bm{\Psi}}^T\bm{U}\right)\right\|_2\\
  &\leq \left\|\widetilde{\bm{\Psi}}\right\|_{2,2}\left\|\bm{\varepsilon}_M\right\|_2+\left\|\bm{U}-\widehat{\bm{\Psi}}\widehat{\bm{\Psi}}^T\bm{U}\right\|_2+
  \left\| \widehat{\bm{\Psi}}\widehat{\bm{\Psi}}^T\bm{U}-\widetilde{\bm{\Psi}}\widetilde{\bm{\Psi}}^T\bm{U}\right\|_2\\
  &\leq \varepsilon_{PCE}+\varepsilon_{ss}+\varepsilon_{rb},
\end{aligned}
\end{equation}
where $ \widehat{\bm{\Psi}}\widehat{\bm{\Psi}}^T$ is the continuous covariance matrix of the random field, which has the form of Eq.~(\ref{conti_version_Cov}), and $\|\cdot\|_{2,2}$ is a matrix norm induced by the $L_2$ norm, e.g.,
$$\left\|\widetilde{\bm{\Psi}}\right\|_{2,2}=\sup_{\bm{y}\neq \bm{0}}\frac{\|\widetilde{\bm{\Psi}}\bm{y}\|_2}{\|\bm{y}\|_2}, \bm{y}\in\mathbb{R}^L.$$
Because $\widetilde{\bm{\Psi}}\widetilde{\bm{\Psi}}^T=\bm{I}_N$, $\left\|\widetilde{\bm{\Psi}}\right\|_{2,2}=1$. Then, we have
\begin{equation}\label{EE}
\begin{aligned}
&\varepsilon_{PCE}=\left\|\widetilde{\bm{\Psi}}\right\|_{2,2}\left\|\bm{\varepsilon}_M\right\|_2=\left\|\bm{\varepsilon}_M\right\|_2,\\
&\varepsilon_{rb}=\left\|\bm{U}-\widehat{\bm{\Psi}}\widehat{\bm{\Psi}}^T\bm{U}\right\|_2\leq \sqrt{\lambda_{L+1}},\\
&\varepsilon_{ss}=\left\|\widehat{\bm{\Psi}}\widehat{\bm{\Psi}}^T\bm{U}-\widetilde{\bm{\Psi}}\widetilde{\bm{\Psi}}^T\bm{U}\right\|_2
\leq \left\|\widehat{\bm{\Psi}}\widehat{\bm{\Psi}}^T-\widetilde{\bm{\Psi}}\widetilde{\bm{\Psi}}^T\right\|_{2,2}\left\|\bm{U}\right\|_2
=\widetilde{C}\|\widetilde{\bm{\Psi}}\widetilde{\bm{\Psi}}^T\|_{2,2}\left\|\bm{U}\right\|_2=\widetilde{C}\left\|\bm{U}\right\|_2,
\end{aligned}
\end{equation}
where $\widetilde{C}$ is a constant related to the method for the snapshot selection, e.g., $\widetilde{C}\sim O\left(N_s^{-1/2}\right)$ for standard Monte Carlo methods \cite{Caflisch1998} and $\widetilde{C}\sim O\left((N_s)^{-k}(\log N_s)^{(d-1)(k+1)}\right)$ for the Smolyak sparse-grid method \cite{Novak1996}, where $k$ is the degree of the polynomials. Therefore, $\varepsilon_{ss}$ mainly depends on the number of snapshots ($N_s$). As discussed in Section 2.1, $\varepsilon_{rb}$ depends on a user-specified tolerance $\varepsilon$; i.e., $\varepsilon_{rb}$ decreases as $\varepsilon$ decreases. $\varepsilon_{PCE}$ depends on the maximum order $p$ and the size of DoE ($N_{\xi}$) \cite{Xiu2010}. In the present study, the snapshots are chosen using a sequential sampling scheme, as shown in Section 2.4. The PCE construction uses a non-intrusive regression-based method, as proposed in \cite{Blatman2011}. Hence, the accuracy of POD--PCE depends on the number of the snapshots ($N_s$), tolerance $\varepsilon$, maximum polynomial order ($p$) and size of DoE ($N_{\xi}$).

%%%%%%%%%%%%%%%%%%%%%%%%%%%%%%%%%%%%%%%%%%%%%%%%%%%%%%%%%%%%%%%%%%%%%
\subsection{Implementation of POD--PCE} \label{Implementation}
In this section, we summarize the overall procedure followed to construct a POD--PCE representation for an $\mathbb{R}^{N}$-valued random field $U(\bm{x};\bm{\xi})$. We first choose a set of snapshots using a sequential sampling method as shown in Section 2.4.
Second, we construct the POD basis using the selected snapshots and obtain a set of reduced basis in the following steps:
\begin{enumerate}
    \item Assemble the snapshot matrix $\bm{\mathcal{U}}=\left[\bm{U}_1, \ldots, \bm{U}_{N_s}\right]$ and solve the eigenvalue problem of $\bm{\mathcal{U}}^T\bm{\mathcal{U}}$ by Eq.~(\ref{EigenPP}) to generate $\lambda_{n_s}$ and $\bm{\nu}_{n_s}$, $n_s=1,2,\ldots,N_s$.
    \item Set $\bm{\psi}_{n_r}=\frac{\bm{\mathcal{U}}\bm{\nu}_{n_r}}{\sqrt{\lambda_{n_r}}}$, $n_r=1,2,\ldots,N_r$.
    \item Reduce the basis to be $\left\{\bm{\psi}_l\right\}_{l=1}^{L}$ such that condition (\ref{Tol}) is satisfied.
    \item Define the reduced POD basis matrix as $\widetilde{\bm{\Psi}}=\left[\bm{\psi}_1,\bm{\psi}_2,\ldots,\bm{\psi}_L\right]$.
\end{enumerate}
Third, a sparse PCE method is utilized to approximate the undetermined coefficients of POD:
\begin{enumerate}
    \item Prepare the DoE as $\left\{\left(\bm{\xi}^{(n_s)}, u_l(\bm{\xi}^{(n_s)})\right),\ n_s=1,2,\dots,N_s\right\}$.
    \item Select a set of orthogonal polynomials  corresponding to the distribution of the random variables.
    \item Construct the PCEs of $u_l(\bm{\xi})$ as Eq.~(\ref{PCE}) by solving the $l_1$ minimization problem (\ref{LS_l1}).
\end{enumerate}
Finally, a POD--PCE representation of $U(\bm{x};\bm{\xi})$ is constructed using Eq.~(\ref{POD_PCE_1}).

All the PCE constructions in this paper are done using the least angle regression algorithm with UQLab \cite{UQLab}.
%It should be noted that the criteria used in choosing the snapshots can be specified as the user requires. For simplicity, we use the selected snapshots and their corresponding random variables as the experiment design in PCE construction. The principles behind choosing the snapshots and the generation of experiment design in PCE are similar, which are to find a subset spanned by the selected snapshots or samples that can be used to represent the original space.

%%%%%%%%%%%%%%%%%%%%%%%%%%%%%%%%%%%%%%%%%
\section{Numerical examples} %Numerical testings
\subsection{Highly irregular Ackley function}
First, we consider a highly irregular Ackley function with three random parameters. The formulation of this function is given by
\begin{align}\label{Ex_1}
 u(\bm{x};\bm{\xi})=&-20(1+0.1\xi_3)\left(\exp\left[-0.2(1+0.1\xi_2)\sqrt{0.5(x^2+y^2)}\right]\right)\\
                    &-\exp(0.5[\cos(2\pi(1+0.1\xi_1)x)+\cos(2\pi(1+0.1\xi_1)y)])+20+e,
\end{align}
where $\bm{\xi}=[\xi_1,\xi_2,\xi_3]$ is a random vector with independent components that are uniformly distributed over $[-1,1]^3$.

\begin{figure}[h!]
\centering
{
  \includegraphics[width=14cm]{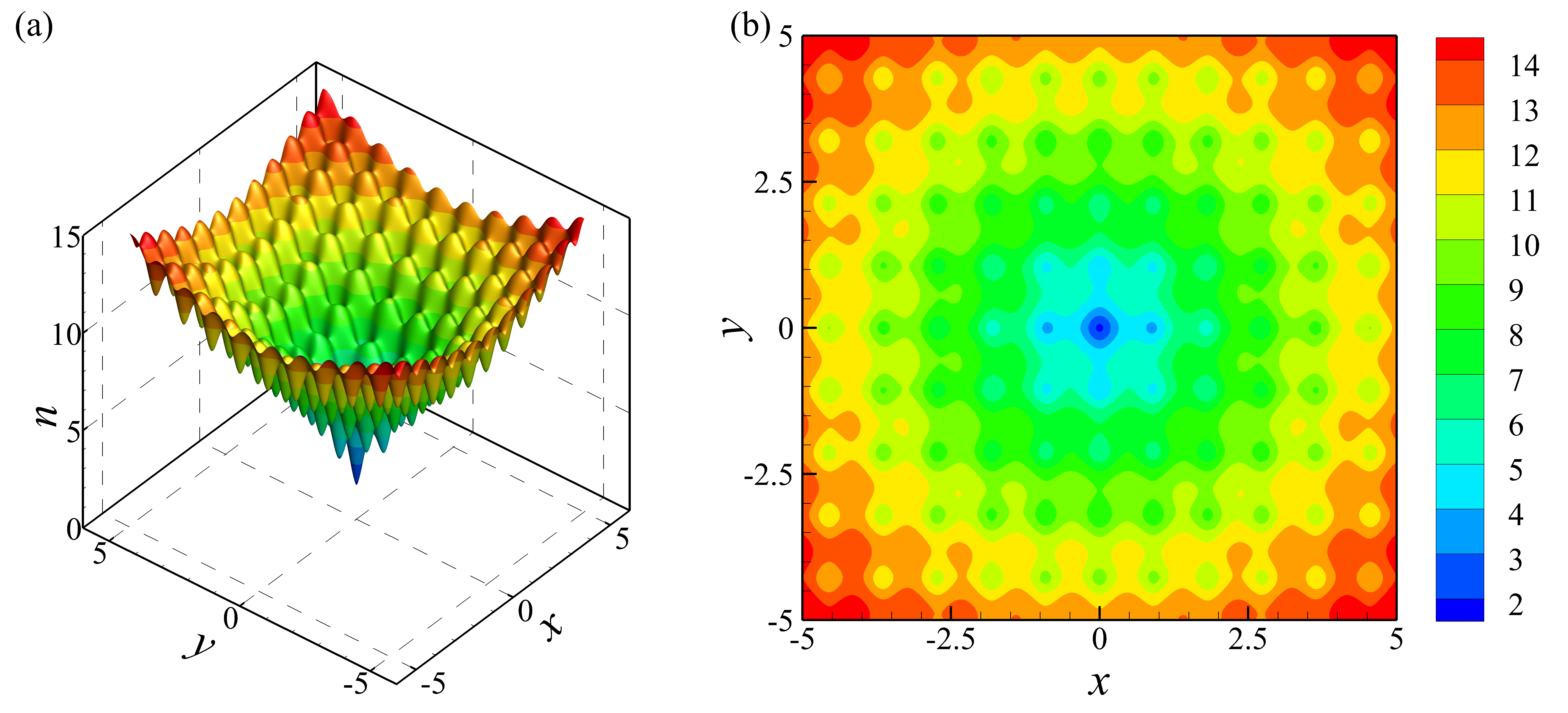}\\[-10pt]
}
\caption{One realization of the random field $u(\bm{x},\bm{\xi})$: (a) 3D plot and (b) contour plot.}\label{Ackley}
\end{figure}

According to the discussion in \cite{Raisee2015}, a discretization with $400\times400$ nodes is required to reproduce the fine-scale structure of the deterministic Ackley function. Hence, such a fine mesh is adopted in the present study for UQ analysis of the stochastic Ackley function. Figure \ref{Ackley} shows the Ackley function on the mesh for a randomly generated parameter vector, where the structure of the function is well depicted.% using a 3D plot and a contour plot.

\begin{figure}[t!]
\centering
\includegraphics[width=9cm]{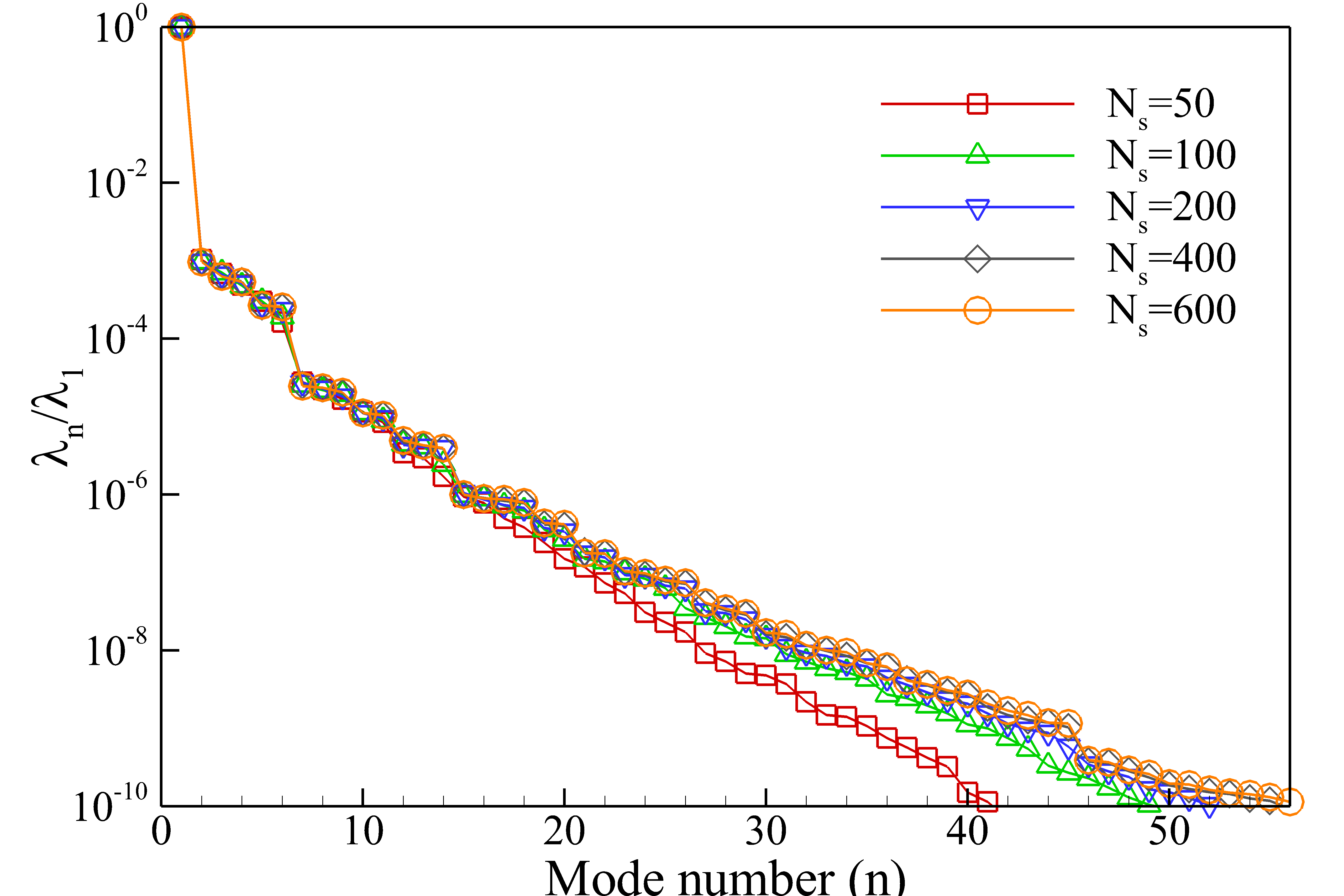}\\[-5pt]
\caption{Normalized eigenvalues calculated using different numbers of snapshots for the stochastic Ackley function.}\label{snapshots_choice}
\end{figure}

As discussed in Section 2, to construct a POD--PCE representation, a proper set of snapshots of a size of $N_s=400$ was first selected by following the procedure described in Section 2.4. % This was because the mean absolute relative error (MARE) of the normalized eigenvalues calculated from the sizes $400$ and $600$ of snapshots is smaller than $0.05$.
Figure \ref{snapshots_choice} shows the normalized eigenvalues calculated using different numbers of snapshots in the snapshot selection procedure. %clearly indicates that eigenvalues calculated from the snapshots with the sizes of $400$ and $600$ are found to be very similar. Here, the eigenvalues are computed and stored in descending order and normalized using the largest one.
We observe that there is a sharp drop in the first seven eigenvalues. This indicates that most of the information in $u(\bm{x; \xi})$ can be captured using the first seven modes. This can be confirmed by Table \ref{Tolerance}, where the result shows that the first seven modes capture $99.99\%$ of the total energy.

With the selected snapshots and corresponding random vectors, we constructed both POD--PCE and full PCE (FPCE) for the output field data of the stochastic Ackley function. Similar to the PCE construction for the POD coefficients in Section 2.2, the full PCE of $u(\bm{x;\xi})$ is constructed as follows:
\begin{equation}\label{OPCE}
  u(\bm{x};\bm{\xi})\approx\sum_{\bm{\alpha}\in \mathcal{A}}c_{\bm{\alpha}}(\bm{x})\Phi_{\bm{\alpha}}(\bm{\xi}).
\end{equation}
The maximum order of the polynomials in the PCE construction parts was set as $p=13$ according to the recommendation in \cite{Raisee2015}. Furthermore, Monte Carlo (MC) simulations were performed using $3\times 10^3$ LHS samples to validate the accuracy of POD--PCE. Given $N_s=400$, we investigated how accurately POD--PCE predicts the statistical quantities by changing the value of $\varepsilon$. The results are presented in Table \ref{Tolerance}, where the number of the selected POD bases (L) and the mean absolute relative errors (MAREs) in mean and variance predictions are presented for different values of $\varepsilon$. The MAREs between POD--PCE and full PCE are defined as follows:
\begin{equation}\label{Tol_errors}
   \varepsilon_{r,\mu}^{FPCE}=\frac{1}{N}\sum_{i=1}^{N}\left|\frac{\mu_{POD-PCE}(i)-\mu_{FPCE}(i)}{\mu_{FPCE}(i)}\right|,\  \varepsilon_{r,var}^{FPCE}=\frac{1}{N}\sum_{i=1}^{N}\left|\frac{var_{POD-PCE}(i)-var_{FPCE}(i)}{var_{FPCE}(i)}\right|.
\end{equation}
Similar definitions can be obtained for the MAREs between POD--PCE and MC; i.e., $\varepsilon_{r,\mu}^{MC}$ and $\varepsilon_{r,var}^{MC}$. The MAREs between POD--PCE and MC (or full PCE) in mean and variance predictions decrease and the number of the selected POD bases increases as the value of $\varepsilon$ decreases from $\varepsilon=10^{-4}$ to $\varepsilon=10^{-8}$. Furthermore, the decreasing feature tends to flat as $\varepsilon$ becomes smaller. In particular, when $\varepsilon\leq 10^{-8}$, the MAREs between POD--PCE and MC are nearly unchanged and those between POD--PCE and full PCE are also very small. This indicates that POD--PCE and PCE have almost the same accuracy in mean and variance predictions. However, it is worth noting that POD-PCE only needs to construct $40$ PCEs for the undetermined coefficients of the selected POD bases for $\varepsilon=10^{-8}$, while full PCE needs to construct $N=160000$ PCEs for all the field data. Clearly, an enormous amount computational cost is reduced using POD--PCE method. In summary, POD--PCE constructed based on $\varepsilon=10^{-8}$ can predict mean and variance with an accuracy similar to that of the classical full PCE economically.

\begin{table}%[]
\small
\centering
\begin{tabular}{ccccccccc}
\hline
 \multirow{2}{*}{\begin{tabular}[c]{@{}c@{}}Tolerance\\ $\varepsilon$\end{tabular}}&  & \multirow{2}{*}{\begin{tabular}[c]{@{}c@{}}Size of selected POD basis\\ $L$\end{tabular}} & & \multicolumn{2}{c}{MARE in mean} & & \multicolumn{2}{c}{MARE in variance}  \\ \cline{1-1}\cline{3-3}\cline{5-6}\cline{8-9}
                          &  &                                  & & $\varepsilon_{r,\mu}^{MC}$                        & $\varepsilon_{r,\mu}^{FPCE}$                     & & $\varepsilon_{r,var}^{MC}$                           & $\varepsilon_{r,var}^{FPCE}$                        \\ \hline
$10^{-4}$                     & &7                                  & &2.4695E-04                 & 2.4117E-04                & &1.8000E-02                   & 1.6200E-02                  \\
$10^{-5}$                  & &13                                 & &6.0376E-05                 & 5.4439E-05                & &5.4000E-03                  & 2.3000E-03                  \\
$10^{-6}$                 & &20                                 & &2.6024E-05                 & 1.7988E-05                & &4.5000E-03                   & 6.7358E-04                  \\
$10^{-7}$                & &29                                 & &1.9915E-05                 & 1.0626E-05                & &4.3000E-03                   & 2.5562E-04                  \\
$10^{-8}$                 & &40                                 & &1.6268E-05                 & 3.2476E-06                & &4.3000E-03                   & 1.1670E-04                  \\
$10^{-9}$               & &52                                 & &1.6278E-05                 & 2.3119E-06                & &4.3000E-03                   & 7.7865E-05                  \\ \hline
\end{tabular}
\caption{Effects of $\varepsilon$ on $L$ and MAREs in mean and variance predictions for $N_s=400$ for the stochastic Ackley function.}\label{Tolerance}
\end{table}
\begin{table}%[]
\small
\centering
\begin{tabular}{ccccccc}
\hline
No. of snapshots& &Size of selected POD basis& & MARE in mean & & MARE in variance\\ \cline{1-1}\cline{3-3}\cline{5-5}\cline{7-7}
$N_s$ & & $L$ & & $\varepsilon_{r,\mu}^{FPCE}$ & & $\varepsilon_{r,var}^{FPCE}$ \\ \hline
50                                                              & & 31                                                          &    & 6.2769E-04                                                                                    & & 3.1000E-02                                                                                        \\
100                                                             & & 37                                                          &    & 2.0937E-04                                                                                   & & 4.0000E-03                                                                                        \\
200                                                             & & 39                                                          &    & 4.0501E-05                                                                                   &  & 9.0066E-03                                                                                        \\
400                                                             & & 40                                                           &   & 3.2476E-06                                                                                    & & 1.1670E-04                                                                                        \\
600                                                             & & 41                                                           &   & 2.6858E-06                                                                                    & & 9.7453E-05\\\hline
\end{tabular}
\caption{Effects of $N_s$ on $L$ and MAREs in mean and variance predictions for $\varepsilon=10^{-8}$ for the stochastic Ackley function. }\label{Ns}
\end{table}

Next, we considered the effects of the number of snapshots ($N_s$) on how accurately POD--PCE predicts the statistical quantities for the tolerance $\varepsilon=10^{-8}$. We are interested in investigating whether the POD--PCE constructed using a small number of snapshots can provide estimations of statistical quantities as reasonable as those obtained using the classical full PCE method. Table \ref{Ns} shows the effects of $N_s$ on the size of the selected POD basis and MAREs in mean and variance predictions for the stochastic Ackley function. The results of the POD--PCE are agreement with those of full PCE as the number of snapshots increases. Furthermore, we observe that even the POD--PCE constructed using $50$ snapshots can predict the mean and variance with reasonable accuracy. However, the error in variance prediction is always larger than that in mean prediction, which implies that higher order moment prediction requires higher accuracy PCE; further, attention should be paid to the high order moment prediction for PCE construction. %This implies that the POD-PCE constructed using a small size of snapshots may give reasonable predictions of statistical quantities in real problems.}

\begin{figure}
\centering
\includegraphics[width=8cm]{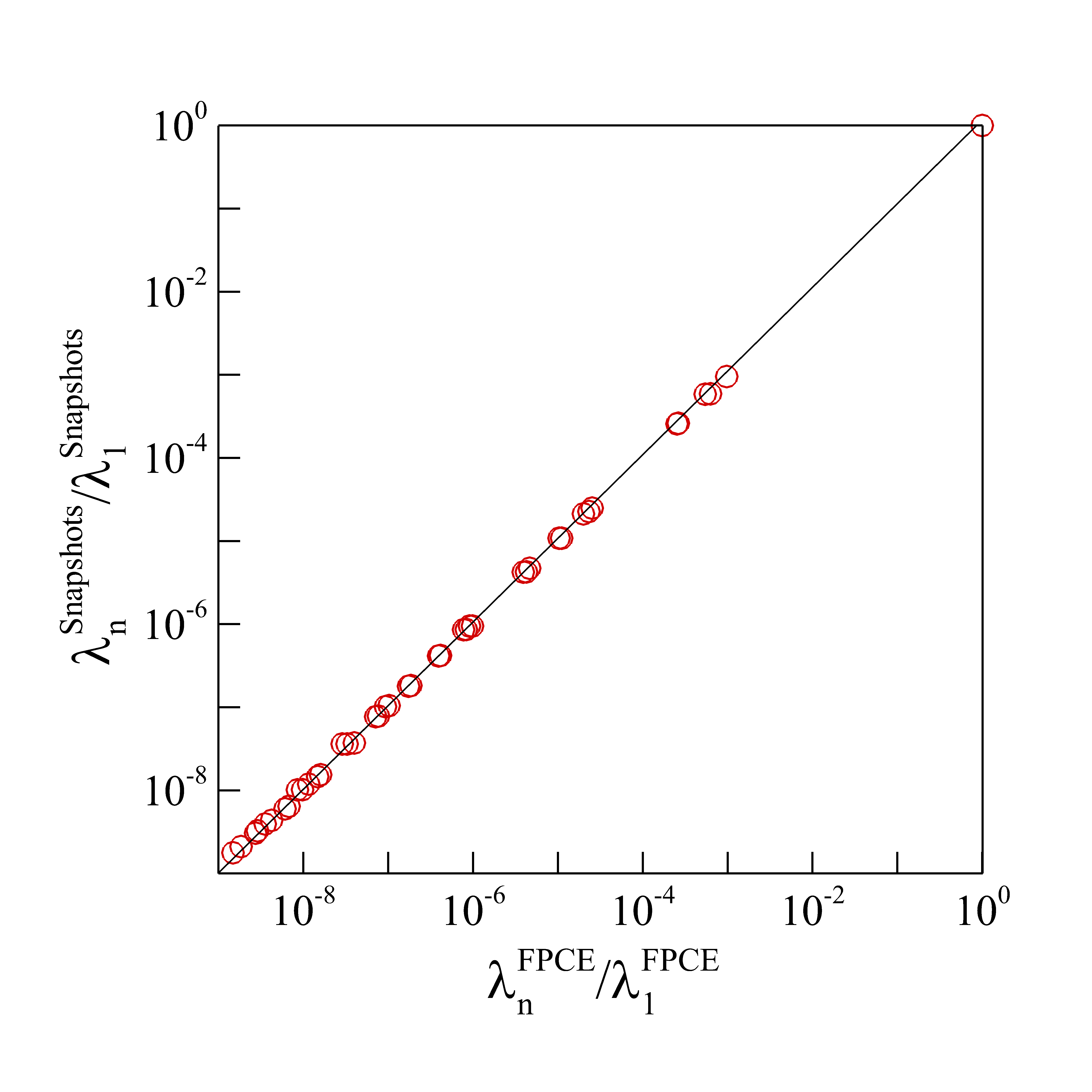}\\[-25pt]
\caption{Comparison of the eigenvalues calculated from the full PCE and the method of snapshots.}\label{eignv_comparison}
\end{figure}

Finally, the efficiency and accuracy of the method of snapshots is validated by the comparison of eigenvalues calculated using full PCE and the method of snapshots. In the method of snapshots, eigenvalues are calculated using $400$ snapshots by Eq.~(\ref{EigenPP}). For the full PCE method, the eigenvalues are obtained by solving the eigenvalue problem of the correlation matrix $\bm{R}=(r_{n_1,n_2})_{N\times N}$ with
\begin{equation}\label{PCE_COV}
  r_{n_1,n_2}=\sum_{m=1}^{M-1}c_{m}(\bm{x}^{(n_1)})c_{m}(\bm{x}^{(n_2)}),\ n_1=1,2,\ldots,N,\ n_2=1,2,\ldots,N,
\end{equation}
where $c_{k}(\bm{x}^{(\cdot)})$ are deterministic coefficients of Eq.~(\ref{OPCE}) corresponding to grid $\bm{x}^{(\cdot)}$. The comparison result is shown in Figure \ref{eignv_comparison}, where we only present eigenvalues corresponding to the modes truncated by the tolerance $\varepsilon=10^{-8}$. For the full PCE method, $42$ bases are selected, which is slightly larger than that in the method of snapshots. However, the comparison shows good agreement for almost all eigenvalues. Only tiny discrepancies exist in the calculation of the smallest eigenvalues. Therefore, the method of snapshots is accurate enough to be used to construct the POD basis compared with the methods discussed in \cite{Doostan2007,Raisee2015,Abraham2018}.

%
%%%%%%%%%%%%%%%%%%%%%%%%%%%%%%%%%%%%%%%%%%%%%%%%%%%%%%%%%%%%%%%%%%%%%%%%%%%%%%
\subsection{Heat-driven cavity flow with a stochastic boundary temperature}

\begin{figure}
\centering
\includegraphics[width=10cm]{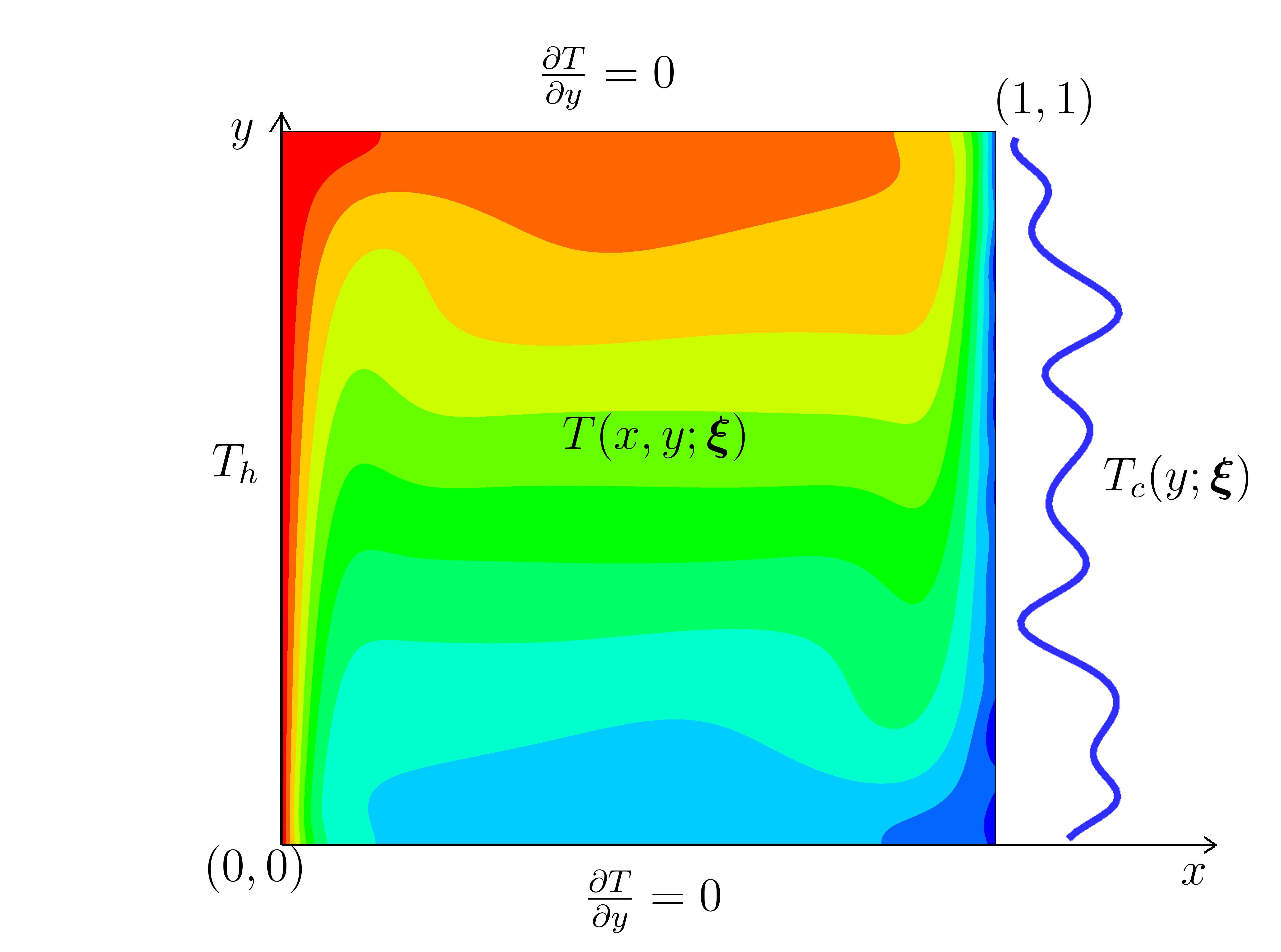}\\[-10pt]
\caption{Illustration of the problem: computational domain, boundary conditions, and a realization of the stochastic temperature field.}\label{schametic}
\end{figure}

To further investigate the performance of the proposed POD-PCE method, we consider a thermally driven cavity flow with stochastic boundary temperatures in a square domain, which is widely used as a numerical example to validate UQ algorithms (see \cite{Maitre2002,Peng2014,Hampton20181} and references therein). Figure \ref{schametic} shows the schematic of the problem with the computational domain, boundary conditions, and a realization of the stochastic temperature field. As described in \cite{Peng2014,Hampton20181}, the boundary temperature $T_h$ on the left vertical wall is uniform, while the boundary temperature $T_c<T_h$ on the right vertical wall is a spatially varying random field.
%As an illustration, several realizations of the stochastic temperature are shown in Figure \ref{schematic}(b).
In terms of dimensionless variables, the time-dependent governing equations for fluids are presented as follows:
\begin{align}
\label{GEmon}
& \f{\pa \bu}{\pa t} + \bu \cdot \nb \bu = -\nb p + \f{Pr}{\sqrt{Ra}} \nb^2 \bu + PrT\bm{e}_y, \\
\label{GEcons}
& \nb \cdot \bu = 0, \\
\label{GEener}
& \f{\pa T}{\pa t} + \nb \cdot (\bu T)= \f{1}{\sqrt{Ra}} \nb^2 T,
\end{align}
where $\bu$, $p$, and $T$ are the velocity vector, pressure, and temperature, respectively. $\bm{e}_y$ denotes the unit normal vector in the vertical dimension. The dimensionless variables employed in the governing equations are defined as follows:
\begin{align}\nonumber
x = \f{\widetilde{x}}{L},~y = \f{\widetilde{y}}{L},~u_i = \f{\widetilde{u}_i}{U_0},~
t = \f{\widetilde{t}}{L/U_0},~p = \f{\widetilde{p}}{\rho U_0^2},~
T = \f{\widetilde{T} - (T_h+\overline{T}_c)/2}{T_h-\overline{T}_c},
\end{align}
\begin{align}\nonumber
Pr = \f{\nu}{\a},~
Ra = \f{g \b (T_h-\overline{T}_c) L^3}{\nu \a},
\end{align}
where $x$ (and $y$), $L=1$, $u_i$, $U_0=\sqrt{g \b L (T_h-\overline{T}_c) \a/\nu}$, $T_h$, and $\overline{T}_c$ respectively represent the Cartesian coordinates, characteristic length, velocity components, characteristic velocity, hot wall temperature, and cold wall mean temperature. In addition, $\a$, $\b=0.5$, $\nu$, $\rho$, and $g$ respectively represent the thermal diffusivity, thermal expansion coefficient, kinematic viscosity, fluid density, and gravitational acceleration. The symbol~~$\widetilde{}$~~indicates a dimensional variable. Here, the Prandtl and Rayleigh numbers are set to $Pr=0.71$ and $Ra=10^6$, respectively.

The temperature on the hot wall is set to a constant value, i.e., $T_h=0.5$. On the cold wall, we consider a boundary temperature with stochastic fluctuations as follows:
\begin{equation}\label{T_cold}
  T_c(x=1, y)=\overline{T}_c+T_c^{'}(y),
\end{equation}
where $\overline{T}_c=-0.5$ is a constant mean temperature and $T_c^{'}(y)$ is the random component, which can be expressed in a truncated Karhunen--Lo\`{e}ve expansion as described in \cite{Peng2014,Hampton20181}:
\begin{equation}\label{KL}
  T_c^{'}(y,\bm{\xi})=\sigma_T\sum_{i=1}^{d}\sqrt{\lambda_i}\psi_i(y)\xi_i,
\end{equation}
where $\xi_i$ are uniform random variables assumed to be independently and uniformly distributed over $[-1,1]$ and $\lambda_i$ and $\psi_i(y)$ are the eigenvalues and eigenfunctions of the exponential correlation function
\begin{equation}\label{correlation}
  C(y_1,y_2)=\sigma_T^2exp\left(-\frac{|y_1-y_2|}{l}\right),
\end{equation}
with $\sigma_T=11/100$ and $l=1/21$. Here, $d=20$ means that only largest eigenvalues and the corresponding eigenfunctions are used to represent the random field $T_c^{'}(y)$ with the random variables $\xi_i$.
\begin{table}%[]
\small
\centering
\begin{tabular}{ccc}
\hline
No. of grid nodes & Averaged Nusselt number ($\overline{Nu}$) & Relative difference \\ \hline
$32\times 32$     & 11.3216                             & $28.2306\%$         \\
$64\times 64$     & 9.2967                              & $5.2970\%$          \\
$128\times 128$   & 8.9193                              & $1.0218\%$          \\
$256\times 256$   & 8.84401                             & $0.1698\%$          \\
$512\times 512$   & 8.82901                             & --           \\ \hline
\end{tabular}
\caption{Grid convergence analysis based on averaged Nusselt number on the hot wall.}\label{GridT}
\end{table}

By Eqs.~(\ref{T_cold}) and (\ref{KL}), deterministic boundary temperatures on the cold wall can be obtained for any random vector $\bm{\xi}=[\xi_1,\ldots,\xi_d]$. Then, computations are performed using the deterministic monolithic projection method developed in \cite{Pan2016,Pan2017,Pan2018}. In this problem, a fine mesh is found by performing grid convergence analysis on the averaged Nusselt number $\overline{Nu}$ on the hot wall for meshes with $32 \times 32$, $64\times 64$, $128\times 128$, $256\times 256$, and $513\times 513$ grid nodes. $\overline{Nu}$ is obtained by integrating the local Nusselt number $Nu = - (\pa T / \pa x) |_{x = 0}$ along the hot wall. The results are present in Table \ref{GridT}. Finally, we chose the mesh with $256\times 256$ grid nodes as the fine mesh because it featured the smallest and most reasonable relative difference from the mesh with $513\times 513$ grid nodes. To investigate the performance of POD--PCE, the output temperature field denoted by $T(\bm{x; \xi})$ was considered in this numerical test.

\begin{figure}
\centering
\includegraphics[width=9cm]{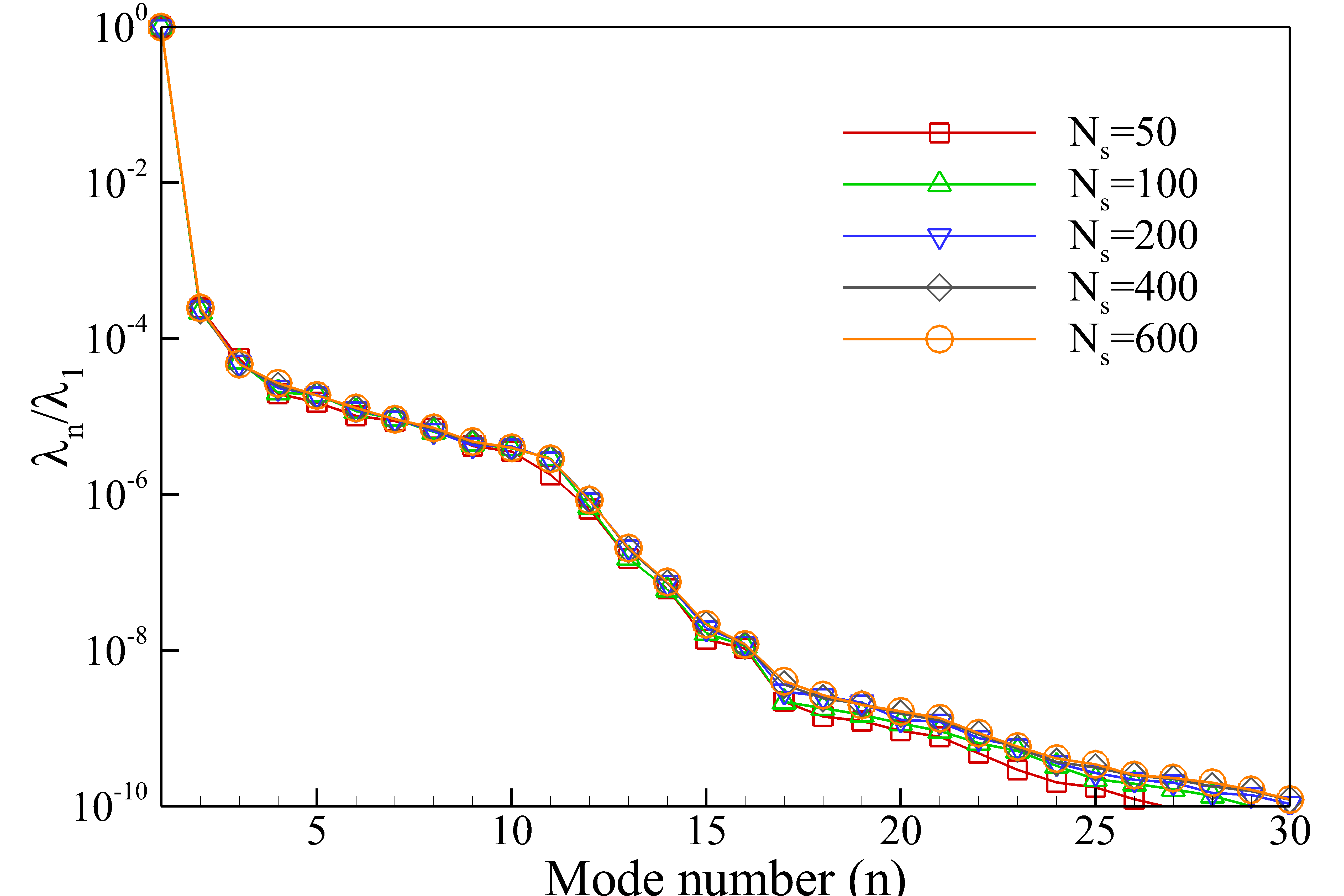}\\[-10pt]
\caption{Eigenvalues calculated for different numbers of snapshots for the stochastic heat-driven cavity flow problem.}\label{snapshots_choice_H}
\end{figure}

Similar to the former example, the size of the snapshots was first chosen as $N_s=400$ using the sequential sampling scheme described in Section 2.4. Figure \ref{snapshots_choice_H} presents all the normalized eigenvalues calculated by the considered snapshots in decreasing order. Compared with the stochastic Ackley function problem, the current calculated eigenvalues decay at a faster rate. Furthermore, we observed a sharp drop in the first three eigenvalues, which means that the first three modes capture most of the information in the temperature field.

\begin{table}%[]
\small
\centering
\begin{tabular}{ccccccc}
\hline
Tolerance& &Size of selected POD basis& & MARE in mean & & MARE in variance\\ \cline{1-1}\cline{3-3}\cline{5-5}\cline{7-7}
$\varepsilon$ & & $L$ & & $\varepsilon_{r,\mu}^{FPCE}$ & & $\varepsilon_{r,var}^{FPCE}$ \\ \hline
$10^{-4}$                     & &3                                 &                 & 8.5751E-05                &                    & 1.6665                 \\
$10^{-5}$                     & &9                                 &                & 9.0676E-05                &                   & 2.4500E-2                  \\
$10^{-6}$                    & &12                                 &                & 9.0842E-05                &                    & 1.3300E-02                  \\
$10^{-7}$                   & &14                                 &                  & 3.0573E-05                &                   & 5.6000E-03                  \\
$10^{-8}$                  & &18                                 &                  & 2.5137E-05                &                    & 5.4000E-03                  \\
$10^{-9}$                  & &28                                 &                 & 2.3924E-05                &                   & 5.4000E-03                  \\ \hline
\end{tabular}
\caption{Effects of $\varepsilon$ on $L$ and MAREs in mean and variance predictions for $N_s=400$ in the stochastic heat-driven cavity flow problem.}\label{H_Tol}
\end{table}
\begin{table}%[]
\small
\centering
\begin{tabular}{ccccccc}
\hline
No. of snapshots& &Size of selected POD basis& & MARE in mean & & MARE in variance\\ \cline{1-1}\cline{3-3}\cline{5-5}\cline{7-7}
$N_s$ & & $L$ & & $\varepsilon_{r,\mu}^{FPCE}$ & & $\varepsilon_{r,var}^{FPCE}$ \\ \hline
50                                                              & & 16                                                          &    & 2.2670E-05                                                                                    & & 7.9990E-01                                                                                        \\
100                                                             & & 17                                                          &    & 2.8925E-05                                                                                   & & 3.9400E-02                                                                                        \\
200                                                             & & 18                                                          &    & 1.5711E-05                                                                                   &  & 7.4000E-03                                                                                        \\
400                                                             & & 18                                                           &   & 2.2587E-06                                                                                    & & 5.4000E-03                                                                                        \\
600                                                             & & 18                                                           &   & 4.1937E-06                                                                                    & & 5.2000E-03\\\hline
\end{tabular}
\caption{Effects of $N_s$ on $L$ and MAREs in mean and variance predictions for $\varepsilon=10^{-8}$ for the stochastic heat-driven cavity flow problem. }\label{H_Ns}
\end{table}
Using $N_s=400$ snapshots, PCEs are constructed with $p=4$ for both POD--PCE and full PCE for $T(\bm{x; \xi})$. The maximum order of polynomials is chosen as that in \cite{Peng2014}, where a good performance of a sparse PCE with $p=4$ is shown in the representation of the stochastic solutions of Eqs.~(\ref{GEmon})--(\ref{GEener}). Further, the fourth-order classical full PCE is constructed using $400$ snapshots and their corresponding random vectors throughout this example. The effects of $\varepsilon$ and $N_s$ on the accuracy of POD--PCE in the prediction of statistical quantities are presented in Tables \ref{H_Tol} and \ref{H_Ns}, respectively. Similar to the previous example, there is a good agreement between POD--PCE results and full PCE results. In particular, the MAREs in mean predictions are very small, even for a large tolerance or fewer snapshots. This is because the first mode is the mean. Thus, if the first mode is selected, we can give a reasonable and accurate estimate of the mean value. For the MAREs in variance predictions, both $\varepsilon$ and $N_s$ have positive effects in that the MAREs decrease with decreasing $\varepsilon$ or increasing $N_s$, and finally, POD--PCE and full PCE obtain similar variance predictions. These indicate that POD--PCE has a similar accuracy to the classical full PCE in the predictions of statistical quantities of $T(\bm{x; \xi})$. Furthermore, it can reasonably predict the statistical quantities with fewer snapshots; e.g., using $N_s=100$ snapshots. In addition, one can find that the MARE of mean prediction increases when $N_s=600$. This is because larger input samples improve the accuracy of POD--PCE. However, the full PCE only used $400$ input samples; the MAREs therefore become larger than for when $N_s=400$. In fact, the mean prediction gradually converges as $N_s$ increases. This is illustrated in Table \ref{PMC}, where $\varepsilon_{r,\mu}^{1000}$ and $\varepsilon_{r,var}^{1000}$ denotes the MAREs computed for $N_s=1000$. %in mean and variance predictions for $\varepsilon=10^{-8}$
\begin{table}%[]
\small
\centering
\begin{tabular}{ccccccc}
\hline
No. of snapshots& &Size of selected POD basis& & MARE in mean & & MARE in variance\\ \cline{1-1}\cline{3-3}\cline{5-5}\cline{7-7}
$N_s$ & & $L$ & & $\varepsilon_{r,\mu}^{1000}$ & & $\varepsilon_{r,var}^{1000}$ \\ \hline
200                                                              & & 18                                                          &    & 3.1563E-05                                                                                    & & 2.4200E-02                                                                                        \\
400                                                             & & 18                                                          &    & 7.1596E-06                                                                                  & & 6.1000E-03                                                                                        \\
600                                                             & & 18                                                          &    & 3.8645E-06                                                                                   &  & 1.5000E-03                                                                                        \\
800                                                             & & 18                                                           &   & 2.6028E-06                                                                                    & & 7.9285E-04                                                                                        \\
1000                                                             & & 18                                                           &   & --                                                                                    & & --\\\hline
\end{tabular}
\caption{MAREs in mean and variance predictions for $\varepsilon=10^{-8}$ for different number of snapshots compared with the case of $N_s=1000$.}\label{PMC}
\end{table}
\begin{figure}[t!]
\centering
{
  \includegraphics[width=1.\textwidth]{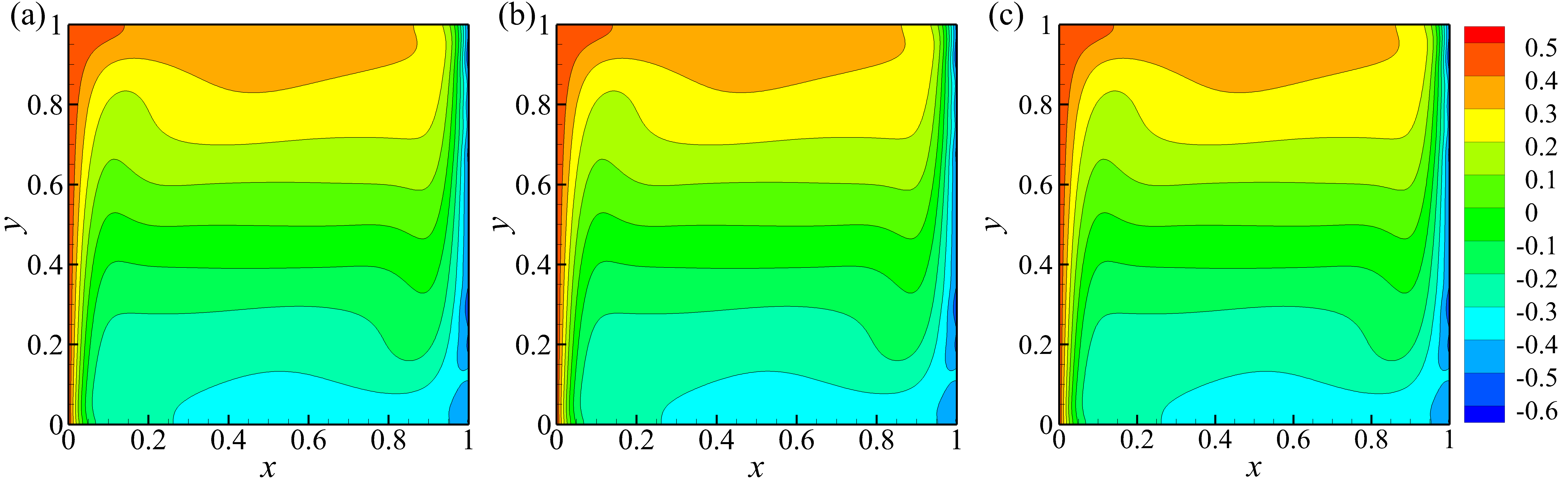}\\[-10pt]
}
\caption{$T(\bm{x;\xi})$ for one randomly generated boundary temperature on the cold wall calculated by: (a) POD--PCE, (b) full PCE, and (c) direct simulation of full order model.}\label{one_prediction}
\end{figure}

To further investigate the accuracy of POD-PCE, we first compared the prediction results of POD--PCE and full PCE for a realization of $T(\bm{x;\xi})$. The results are shown in Figure \ref{one_prediction} with the result directly simulated from Eqs.~(\ref{GEmon})--(\ref{GEener}). We observe that the contours of temperature fields are similar, and it is difficult to distinguish differences among these contour plots. This indicates that both full PCE and POD--PCE can give an accurate prediction of $T(\bm{x; \xi})$ and the errors in the full PCE and POD--PCE predictions are comparable. However, we guess that a few larger errors in the predictions of POD--PCE and full PCE may exist near the right cold wall due to the uncertain temperature boundary condition.

\begin{figure}[t!]
\centering
{
  \includegraphics[width=14cm]{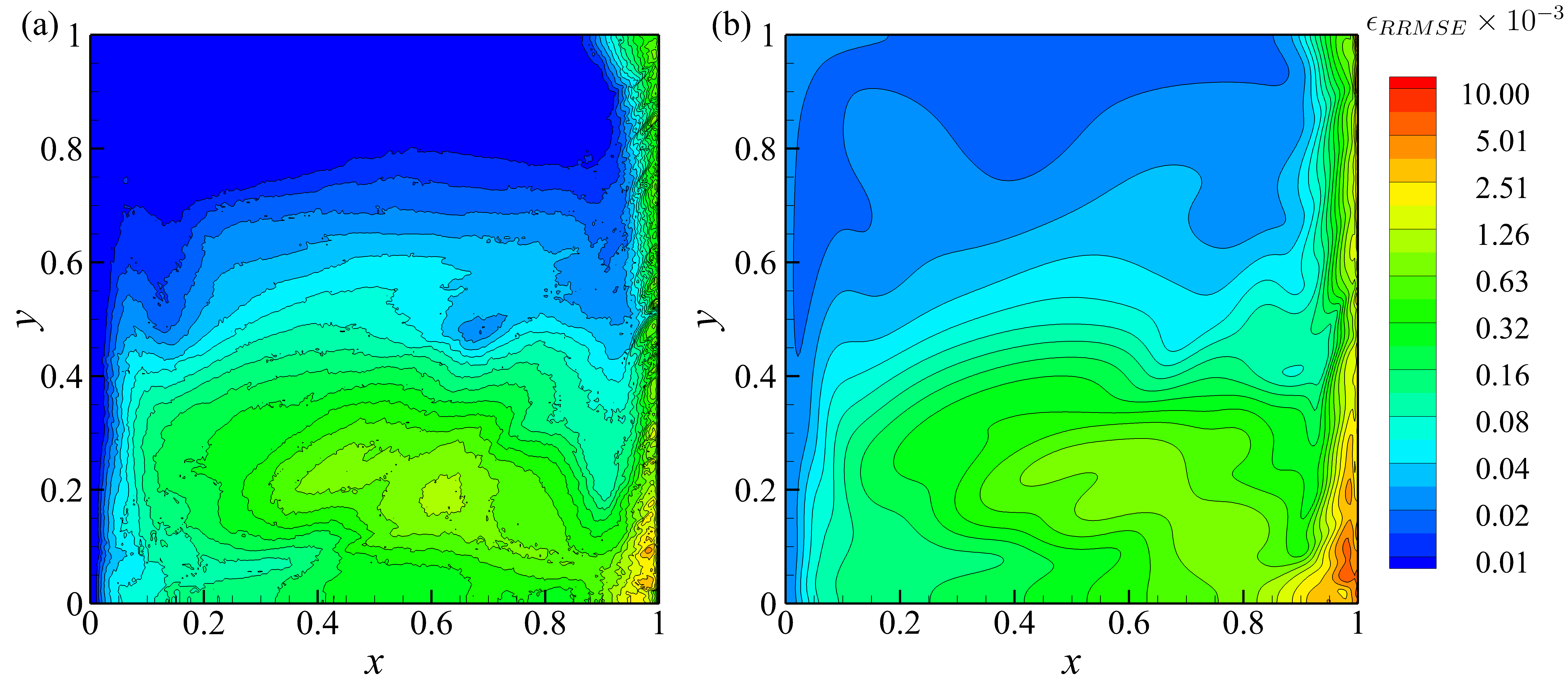}\\[-10pt]
}
\caption{Distributions of RRMSE for the predictions of $T(\bm{x;\xi})$ in : (a) full PCE and (b) POD--PCE. Two cases are performed based on $100$ randomly generated boundary temperatures on the cold wall.}\label{RRMSE_field}
\end{figure}

\begin{figure}[t!]
\centering
{
  \includegraphics[width=14cm]{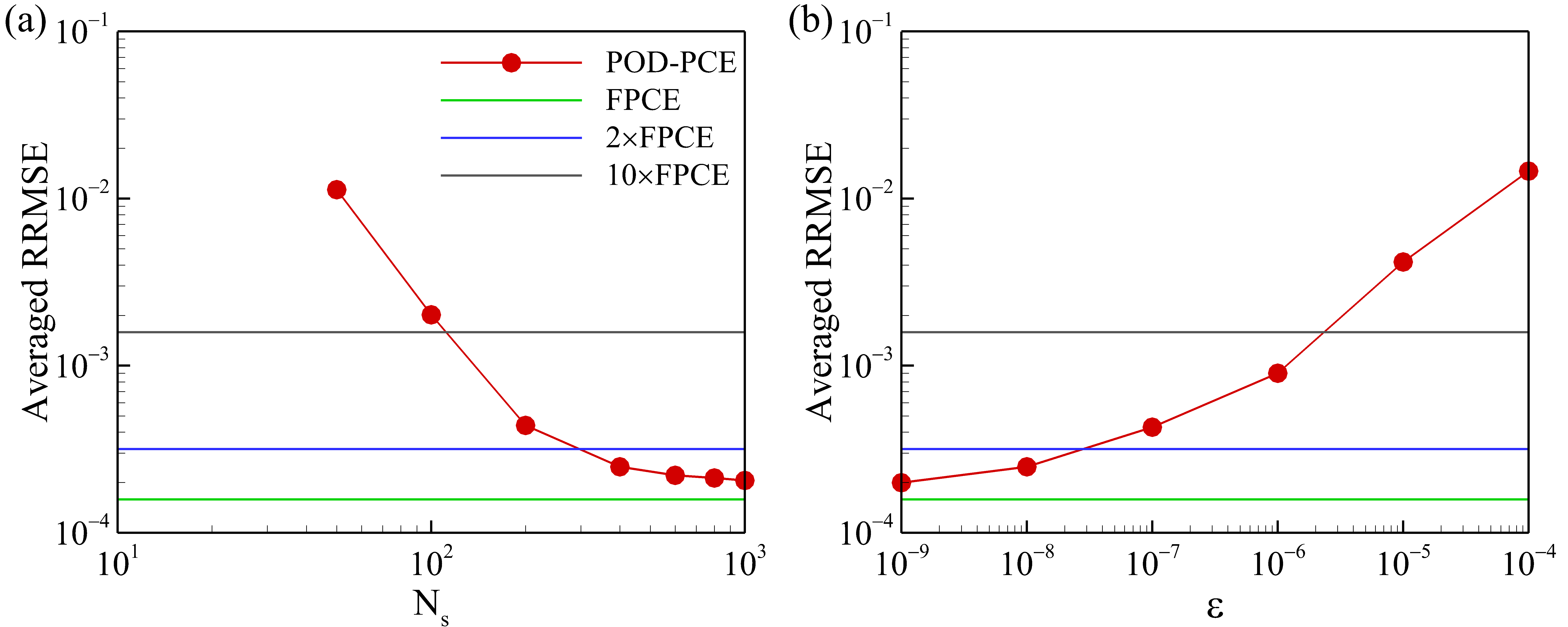}\\[-10pt]
}
\caption{Averaged RRMSE of $T(\bm{x;\xi})$ predications using the POD--PCE method as a function of (a) the number of snapshots and (b) size of selected POD basis. All results are calculated based on $100$ randomly generated boundary temperatures on the cold wall. }\label{ARRMSE_ns_L}
\end{figure}

Next, we measured the error in the predictions of POD--PCE and full PCE by relative root mean-squared error (RRMSE), defined by
\begin{equation}\label{RRMSE}
  \varepsilon_{RRMSE}(\bm{x})=\frac{\sqrt{\frac{1}{N_s}\sum_{n_s=1}^{N_s}\left(T_{LF}(\bm{x};\bm{\xi}^{(n_s)})-T_{HF}
  (\bm{x};\bm{\xi}^{(n_s)})\right)^2}}{\sqrt{\frac{1}{N_s}\sum_{n_s=1}^{N_s}T_{HF}(\bm{x};\bm{\xi}^{(n_s)})^2}},
\end{equation}
where $HF$ denotes the high-fidelity result directly simulated from the full order model and $LF$ represents the low-fidelity result estimated from POD--PCE or full PCE. The results are shown in Figure \ref{RRMSE_field}, where we compare the distributions of RRMSE for the predictions of $T(\bm{x;\xi})$ in POD--PCE and full PCE. The RRMSEs are computed based on $100$ realizations of $T(\bm{x; \xi})$ calculated via direct simulation, POD--PCE, and full PCE for $100$ randomly generated input variable vectors $\{\bm{\xi}^{(k)}\}_{k=1}^{100}$. Large RRMSEs are mainly distributed near the right cold wall, especially at the right bottom angle of the computational domain. Furthermore, larger RRMSEs exist in the predictions of POD--PCE compared with those in full PCE. This is because of the error introduced by the truncated POD in POD--PCE method. In fact, these larger RRMSEs may tend to the values of the corresponding RRMSEs of the full PCE when we decrease the selection tolerance, which can be illustrated in Figure \ref{ARRMSE_ns_L}(b). In addition, we can find that the values of RRMSE of POD--PCE and full PCE are comparable in the predictions of $T(\bm{x;\xi})$. This indicates that the POD--PCE has an accuracy comparable to that of full PCE in the prediction of $T(\bm{x; \xi})$.
%%\XS{Need revision}\\\
Finally, we investigate the effects of the number of snapshots ($N_s$) and the tolerance ($\varepsilon$) of POD basis selection on the accuracy of POD--PCE using the averaged RRMSEs over the temperature field. By Eq.~(\ref{RRMSE}), the averaged RRMSEs are given by
\begin{equation}\label{ARRMSE}
  \hat{\varepsilon}_{RRMSE}=\frac{1}{N}\sum_{n=1}^{N} \varepsilon_{RMSE}(\bm{x}^{(n)}).
\end{equation}
For a fixed tolerance $\varepsilon=10^{-8}$, the averaged RRMSEs of $T(\bm{x;\xi})$ predications in the POD--PCE method are calculated for the aforementioned $100$ realizations for different $N_s$. The results are shown in Figure \ref{ARRMSE_ns_L}(a). The error decreases with the increasing number of snapshots. However, the error decreasing rate becomes smaller simultaneously. In general, when $100<N_s<400$, POD--PCE has a comparable averaged RRMSE to that of full PCE and when $N_s>400$, a similar averaged RRMSE to that of full PCE can be obtained. The difference between two average RRMSEs at $N_s=400$ is based on POD truncation. Therefore, the POD truncation tolerance ($\varepsilon$) should be decreased to improve the accuracy of POD--PCE. This is studied in Figure \ref{ARRMSE_ns_L}(b), where the averaged RRMSEs are computed for a fixed number of snapshots; i.e., $N_s=400$. Clearly, we observe that the error decreases as $\varepsilon$ decreases. Furthermore, we find that, when $\varepsilon\leq 10^{-8}$, POD--PCE can have similar averaged RRMSE to that of full PCE. In summary, if we use the same expansion order in the PCE construction parts for both POD--PCE and full PCE, the accuracy of POD--PCE depends on $N_s$ and $\varepsilon$. By accurately choosing the number of snapshots and tolerance value, the proposed POD--PCE method can deliver accuracy similar to that of the classical full PCE method. %In this example, POD-PCE with $N_s=400$ and $\varepsilon=10^{-8}$ is sufficient to obtain similar accuracy to that of the full PCE method in the prediction of $T(\bm{x,\xi})$ and its statistical quantities.
\subsection{Two-dimensional heat diffusion with stochastic conductivity}%
\begin{figure}[t!]
\centering
{
  \includegraphics[width=14cm]{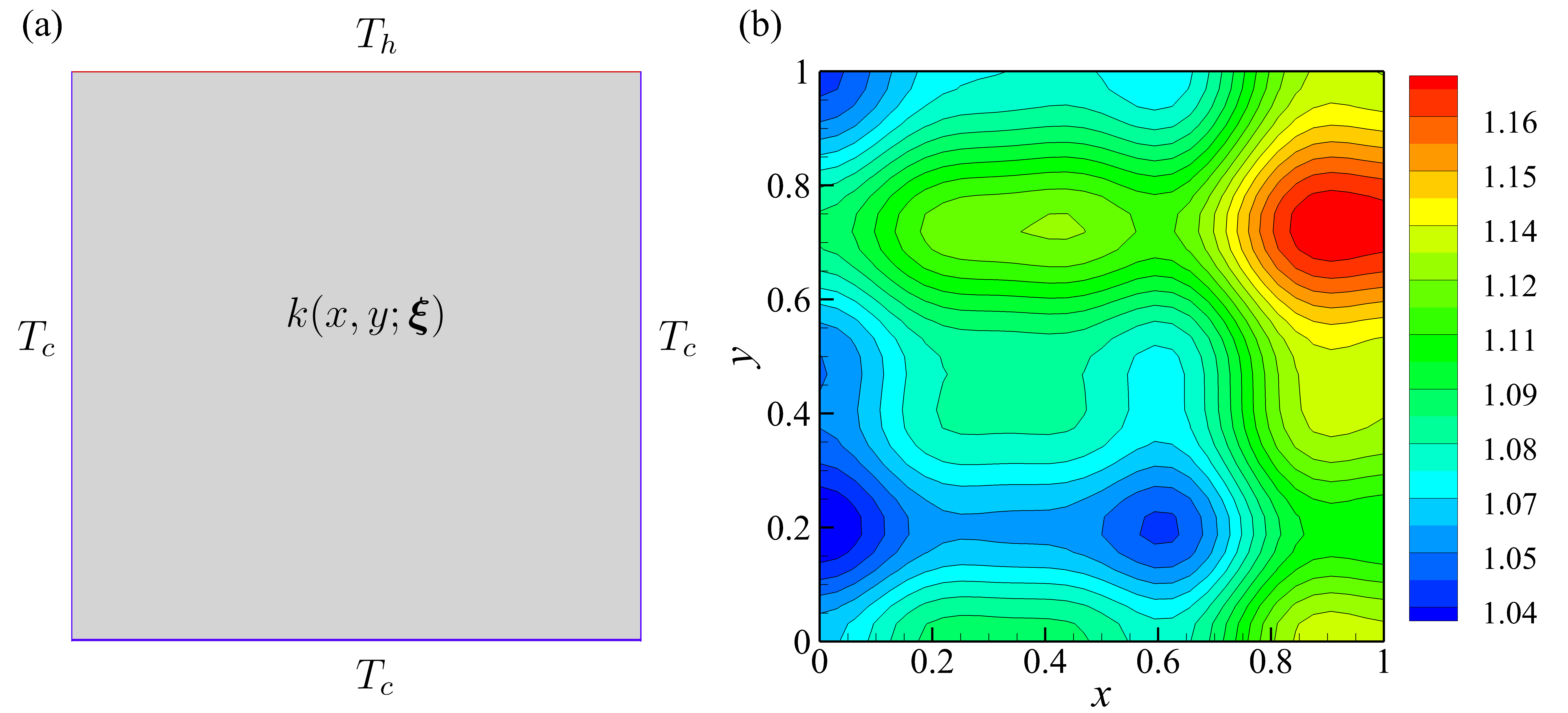}\\[-10pt]
}
\caption{Illustration of the problem: (a) the schematic of the computational domain and boundary conditions, and (b) one realization of the thermal conductivity field.}\label{HD_domain}
\end{figure}

In this section, we consider a two-dimensional (2D) steady-state heat diffusion in a square plate. The governing equation is described as follows:
\begin{equation}\label{HD_governingEq}
  \frac{\partial}{\partial x}\left(k(x,y;\bm{\xi})\frac{\partial T}{\partial x}\right)+\frac{\partial}{\partial y}\left(k(x,y;\bm{\xi})\frac{\partial T}{\partial y}\right)=0,
\end{equation}
where $k(x,y;\bm{\xi})$ is assumed to be a 2D random field defined as a function of the spatial vector $\bm{x}=(x,y)$ as well as the random vector $\bm{\xi}$. The schematic of the computational domain, boundary conditions, and a realization of the thermal conductivity are shown in Figure \ref{HD_domain}. The top boundary of the plate is at a high temperature $T_h$ and other boundaries of the plate are at a low temperature $T_c$. The thermal conductivity $k(\bm{x};\bm{\xi})$ is defined over $\mathcal{D}=[0,1]^2$ with known mean $\bar{k}(\bm{x})$ and covariance function:
\begin{equation}\label{HD_covariance}
  R(\bm{x_1,x_2})=\sigma^2e^{-|x_1-x_2|/l_1-|y_1-y_2|/l_2},
\end{equation}
where $\sigma$ is standard deviation on the conductivity and $l_1$ and $l_2$ are the correlation lengths in $x$ and $y$ directions, respectively.

According to KL expansion, we can approximate $k(\bm{x;\xi})$ with $d$ inputs random variables as follows
\begin{equation}\label{HD_KL}
  k(\bm{x;\xi})\approx \bar{k}(\bm{x})+\sigma\sum_{n=1}^d\left[\sqrt{\lambda_n}\phi_n(\bm{x})\xi_n\right],
\end{equation}
where $d$ is the truncation term and $\lambda_n$ and $\phi_n(\bm{x})$ are respectively eigenvalues and eigenfunctions obtained by solving the following 2D integral equation:
\begin{equation}\label{HD_EignvP}
  \int_{\mathcal{D}} R(\bm{x_1,x_2})\phi_n(\bm{x_2})d\bm{x_2}=\lambda_n\phi_n(\bm{x_1}).
\end{equation}
For any given random vector $\bm{\xi}$, $k(\bm{x;\xi})$ can be approximated by Eq.~(\ref{HD_KL}). The deterministic computation of Eq.~(\ref{HD_governingEq}) is then solved using a centered finite difference scheme with Jacobi iterative method on a uniform grid. In the present study, the boundary temperatures are set to $T_h=1$ and $T_c=0$, the thermal conductivity mean and standard deviation are respectively set to $\bar{k}(\bm{x})=1$ and $\sigma=0.2$, and the correlation lengths are assumed to be $l_1=l_2=6$. In addition, the number of grid nodes is set to $N$ and the stopping criterion in the deterministic solver is set to be $10^{-10}$.
\begin{table}%[]
\small
\centering
\begin{tabular}{ccc}
\hline
No. of grid nodes ($N$) & Average value of $T(\bm{x})$ ($\frac{1}{N}\sum T(x,y)$) & Relative difference \\ \hline
$9\times 9$     & 0.2344                             & $6.16\%$         \\
$17\times 17$   & 0.2461                             & $1.47\%$          \\
$33\times 33$   & 0.2490                             & $0.29\%$          \\
$65\times 65$   & 0.2498                             & --          \\ \hline
\end{tabular}
\caption{Grid convergence analysis based on the average value of temperature field on different meshes.}\label{HD_mesh}
\end{table}

Table \ref{HD_mesh} shows the result of the grid convergence analysis based on the average values of temperature field on different meshes. It is clearly observed that the mesh with grid nodes $N=33\times 33$ is fine enough to be used to simulate the present heat diffusion problem. Thus, the following analyses are performed based on this mesh.
\begin{figure}[t!]
\centering
{
  \includegraphics[width=1.\textwidth]{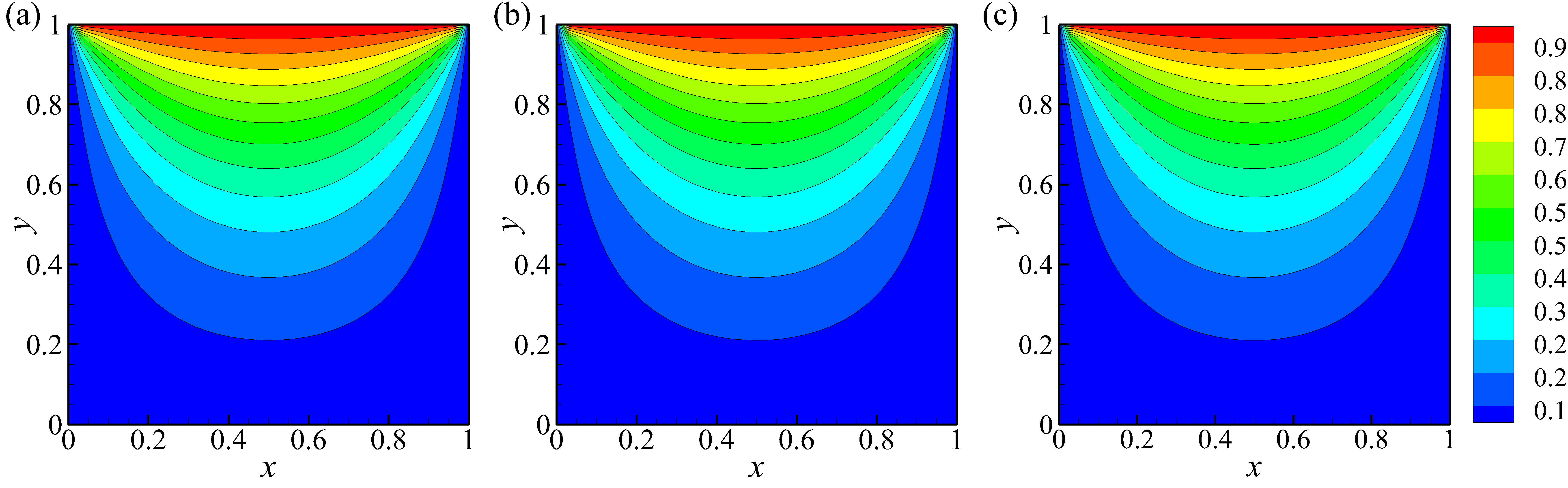}\\[-10pt]
}
\caption{$T(\bm{x;\xi})$ for one randomly generated thermal conductivity by: (a) POD--PCE, (b) full PCE and (c) direct simulation of full order model.}\label{HC_one_prediction}
\end{figure}
In this example, we focused on investigating the accuracy of the proposed POD--PCE method in the prediction of $T(\bm{x;\xi})$. Similar to the previous examples, $N_s=400$ snapshots was first chosen using the selection procedure in Section 2.4. The POD truncation criterion was defined as $\varepsilon=10^{-9}$ and $L=17$ POD bases were then selected based on this criterion. Next, we constructed the PCEs for both POD--PCE and full PCE methods using the DoE of the size $N_s=400$ as described in Section 2.6. Furthermore, all the PCEs are built using the third-order Legendre polynomials ($p=3$).

\begin{figure}[t!]
\centering
{
  \includegraphics[width=14cm]{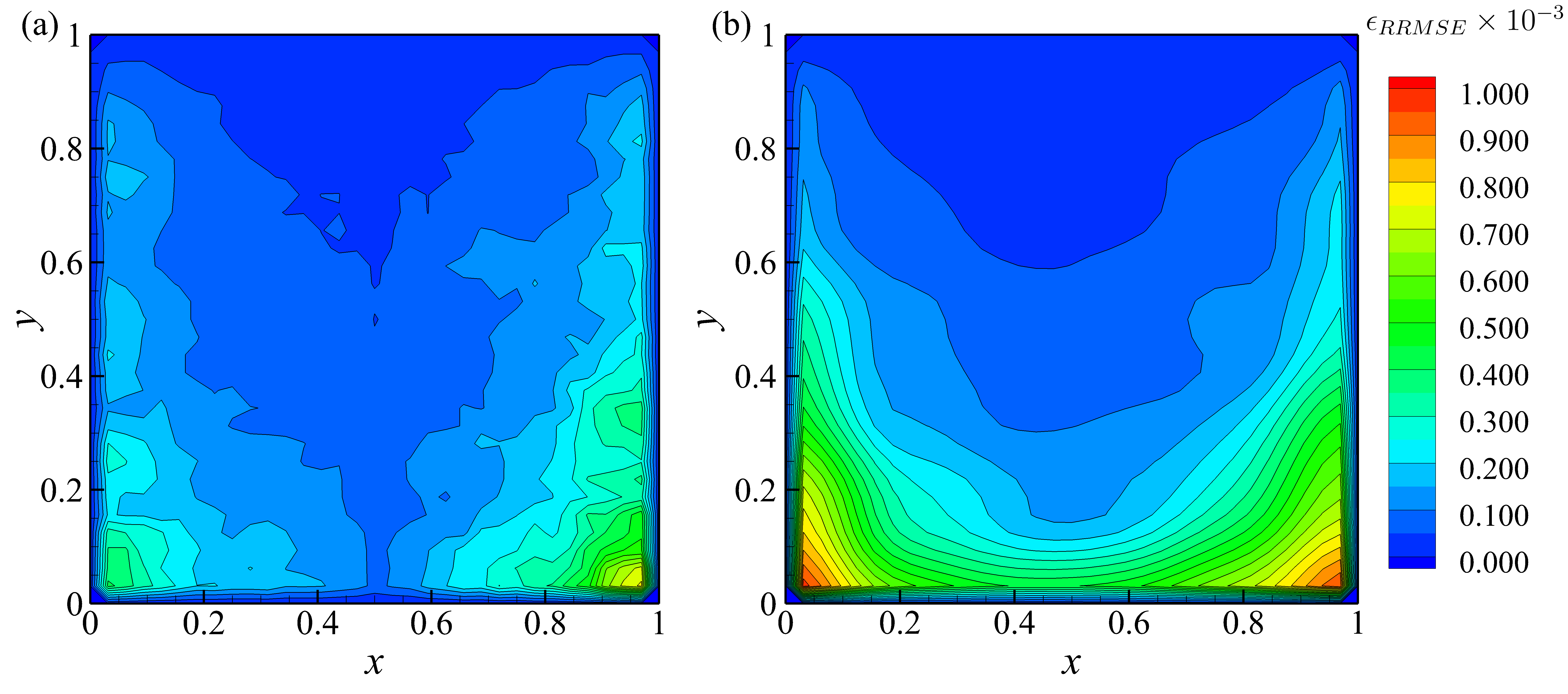}\\[-10pt]
}
\caption{Distributions of RRMSE for the predictions of $T(\bm{x;\xi})$ in : (a) full PCE and (b) POD--PCE. Two cases are performed based on $100$ randomly generated thermal conductivity field.}\label{HC_RRMSE_field}
\end{figure}

\begin{figure}[t!]
\centering
{
  \includegraphics[width=14cm]{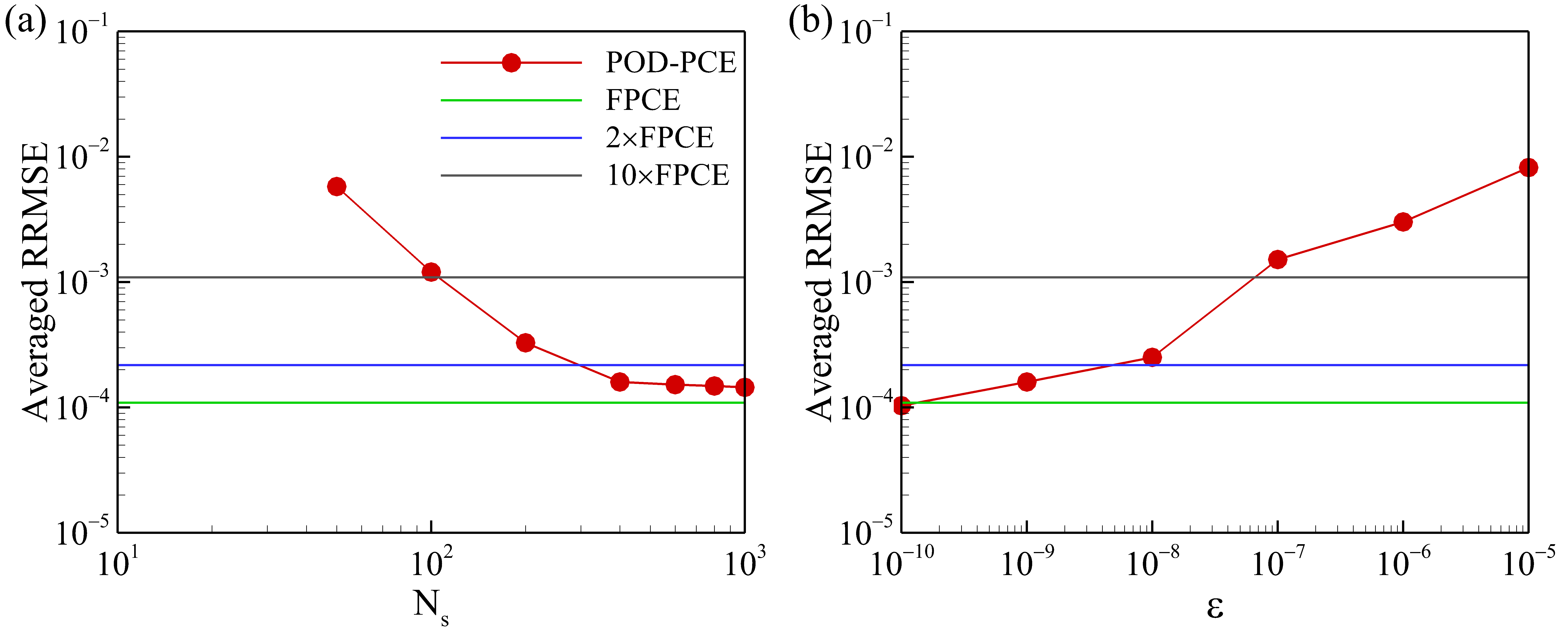}\\[-10pt]
}
\caption{Averaged RRMSE of $T(\bm{x;\xi})$ predications using the POD--PCE method as a function of (a) the number of snapshots and (b) size of selected POD basis. All results are calculated based on $100$ randomly generated thermal conductivity fields. }\label{HC_ARRMSE_ns_L}
\end{figure}

For a randomly generated thermal conductivity, we estimated $T(\bm{x;\xi})$ using POD--PCE and full PCE, respectively. The results are shown in Figure \ref{HC_one_prediction}, compared with the results directly simulated using the full order model. In general, we observe a good agreement among these results. This indicates that the classical full PCE and POD--PCE both provide reasonable estimations for $T(\bm{x;\xi})$. To further investigate the prediction accuracy of the proposed POD--PCE method, we measured the error in the predictions of POD-PCE and full PCE by the RRMSE as defined in Eq. (\ref{RRMSE}). Figure \ref{HC_RRMSE_field} shows the distributions of RRMSE for the predictions of $T(\bm{x;\xi})$ in full PCE and POD--PCE. The RRMSEs are calculated based on $100$ randomly generated thermal conductivity fields. Similar to heat-driven cavity flow, the distributions of the two RRMSE fields are similar, whereas more larger values are found in the RRMSE field for the POD--PCE method, especially at the bottom corners of the domain. This implies that two methods have similar accuracy. However, the performance of the full PCE is slightly better than that of the POD--PCE method. This is induced by the errors coming from the POD construction. We can improve the accuracy of the POD--PCE method by increasing the number of snapshots ($N_s$) or decreasing the POD truncation tolerance ($\varepsilon$). This is illustrated in Figure \ref{HC_ARRMSE_ns_L}, where the averaged RRMSEs of $T(\bm{x;\xi})$ predictions using POD--PCE method for different number of snapshots and truncation tolerances are shown. The results in Figure \ref{HC_ARRMSE_ns_L}(a) are calculated based on a fixed $\varepsilon=10^{-9}$, while the results in Figure  \ref{HC_ARRMSE_ns_L}(b) are calculated based on a fixed $N_s=400$. From the figure, we observe that the error is continuously decreasing for a fixed $\varepsilon=10^{-9}$ when $N_s>400$ though the decreasing rate is very small. In particular, for $N_s=400$, we find that the averaged RRMSE of POD--PCE is similar to that of full PCE when $\varepsilon=10^{-10}$. This re-emphasizes that POD--PCE can reach a similar accuracy as full PCE when we use the same size DoE in both methods.
% Clearly, we observe that the averaged RRMSE decreases as $n_s$ increases or as $\varepsilon$ decreases. In particular, when $n_s\geq 200$, POD-PCE has a comparable accuracy to that of full PCE and a similar accuracy is obtained when $n_s\geq 400$. Further, it seems like that the POD-PCE may has the same accuracy to that of full PCE as the snapshot size continue to increase. When snapshot size is fixed (e.g., $n_s=400$), a comparable accuracy of POD-PCE to that of full PCE can be obtained when $\varepsilon\leq 10^{-6}$ and a similar accuracy is obtained when $\varepsilon\leq 10^{-8}$. Also the same accuracy to that of full PCE may be obtained if we continue to decrease $\varepsilon$. These indicate that POD-PCE can have a similar accuracy to that of the classical full PCE. Furthermore, the same accuracy may be obtained to that of full PCE if we increase the size of snapshots or decrease the selection tolerance of the POD basis.
%%

Furthermore, we compared our method with the reduced-order modeling method (physical-base ROM-II) proposed in \cite{Audouze2009} from the view of computational cost using the above three examples. The results are presented in~\ref{appedix-a}. It is found that the physical-based ROM-II is usually faster than POD--PCE under the same DoE. However, the accuracy of the physical-based ROM-II is worse than that of the POD--PCE. To obtain a similar accuracy as the POD-PCE method, more simulations of the full order model are needed leading to higher computational cost compared with the POD--PCE method. In addition, the proposed POD--PCE method is much easier on the estimation of statistical quantities than the physical-base ROM-II method owing to its orthogonal property. Thus, the proposed POD--PCE is more efficient than the ROM-II method in the prediction of model outputs.

%%%%%%%%%%%%%%%%%%%%%%%%%%%%%%%%%%%%%%%%%
\section{Conclusion}
We developed a non-intrusive reduced-order modeling method for the stochastic representations in an uncertainty quantification analysis of problems with high-dimensional physical and random spaces. The method is based on a model reduction technique POD and an efficient stochastic spectral method PCE. POD is used to extract a set of reduced basis functions from a data ensemble collected from the full order model; i.e., snapshots. PCE is used to approximate the undetermined coefficients of the reduced basis. In particular, a non-intrusive sparse PCE is used to deal with highly random dimensionality problems. The proposed method inherits advantages of both POD and PCE methods. It can provide a reduced-order representation over the physical space as well as efficiently estimate the statistical quantities without increased computational cost. Three examples were tested to investigate the performance of the proposed method, i.e., an analytic highly irregular Ackley function with three random parameters, a heat-driven cavity flow with a stochastic boundary temperature, and a 2D heat diffusion with stochastic conductivity. The results show that the proposed method can reduce the model order from properly chosen snapshots at a reasonable accuracy. Furthermore, under the same parameter setting for the PCE construction, the proposed method can predict the model response and its statistical quantities with an accuracy similar to that of the classical full PCE method economically. In addition, the proposed method is more efficient compared with the existing radial basis function-based ROM-II proposed in \cite{Audouze2009}. These imply great potential for the proposed method applied in an uncertainty quantification analysis.
%%%%%%%%%%%%%%%%%%%%%%%%%%%%%%%%%%%%%%%%%
\section*{Acknowledgements}
This work was supported by the National Research Foundation of Korea(NRF) grant funded by the Korea government(Ministry of Science and ICT) (NRF-2017R1E1A1A0-3070161 and NRF-20151009350).
%%%%%%%%%%%%%%%%%%%%%%%%%%%%%%%%%%%%%%%%%
%%%%%%%%%%%%%%%%%%%%%%%%%%%%%%%%%%%%%%%%%

\appendix
\section{Comparison between POD--PCE and ROM-II}\label{appedix-a}

\begin{figure}[h]
\centering
{
  \includegraphics[width=1.\textwidth]{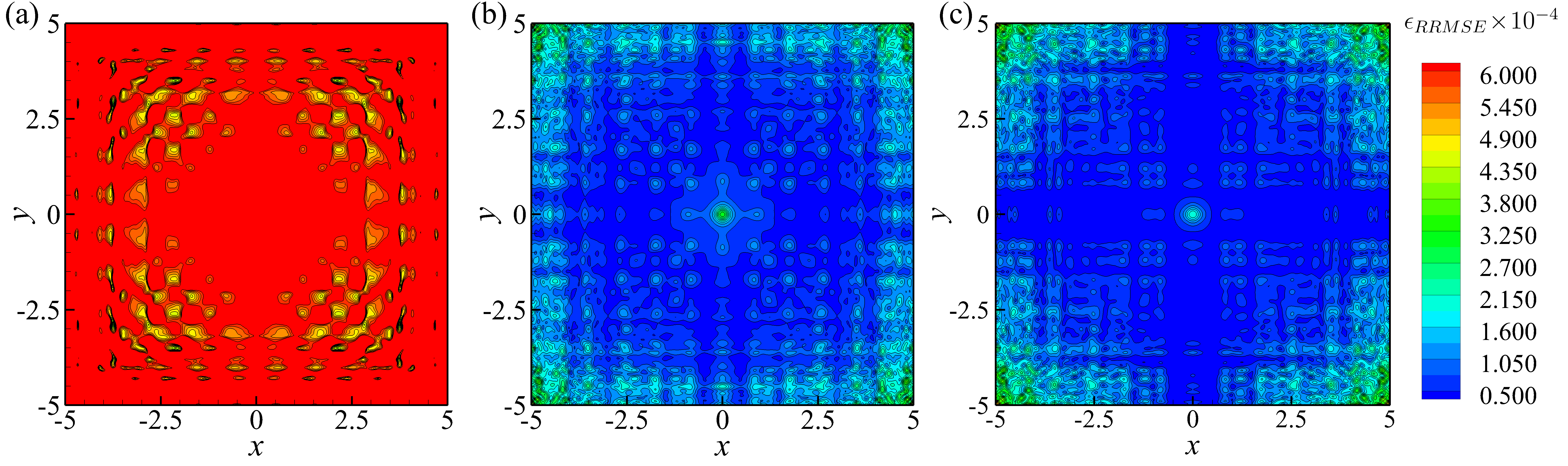}\\[-10pt]
}
\caption{RRMSEs of $T(\bm{x};\bm{\xi})$ for 100 randomly generated parameters from (a) ROM-II (400), (b) ROM-II (10000), and (c) POD--PCE (400).}\label{Comparison_AK}
\end{figure}

\begin{figure}[t!]
\centering
{
  \includegraphics[width=1.\textwidth]{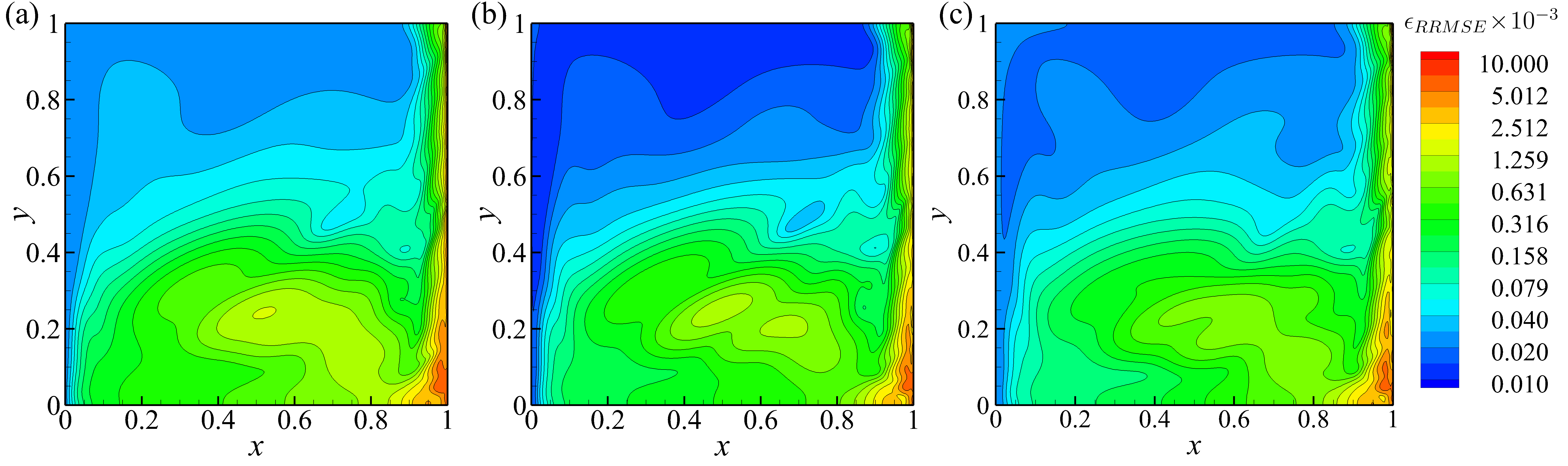}\\[-10pt]
}
\caption{RRMSEs of $T(\bm{x};\bm{\xi})$ for 100 randomly generated boundary temperatures on the cold wall from (a) ROM-II (400), (b) ROM-II (1000), and (c) POD--PCE (400).}\label{Comparison_HD}
\end{figure}

\begin{figure}[t]
\centering
{
  \includegraphics[width=1.\textwidth]{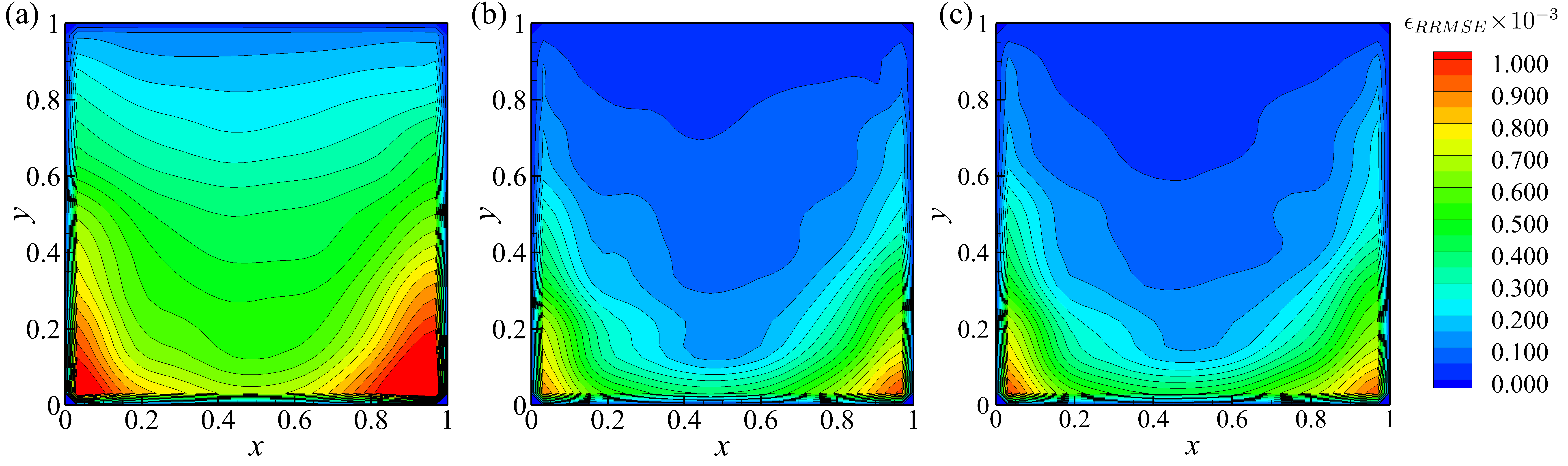}\\[-10pt]
}
\caption{RRMSEs of $T(\bm{x};\bm{\xi})$ for 100 randomly generated conductivity fields from (a) ROM-II (400), (b) ROM-II (2800), and (c) POD--PCE (400).}\label{Comparison_HC}
\end{figure}

This section provides a comparison between the proposed POD--PCE method and ROM-II method proposed in \cite{Audouze2009}. To this end, three examples in Section 3 are tested. The RRMSE defined in Eq.~(\ref{RRMSE}) is estimated as the measure of comparison. The results are shown in Tables \ref{Ackley_cost}--\ref{heat-diffusion} and Figures \ref{Comparison_AK}--\ref{Comparison_HC}.
%\subsubsection{Highly irregular Ackley function}

\begin{table}[h]
\small
\centering
\begin{tabular}{ccccccccc}
\hline
\multirow{2}{*}{Method} &\multirow{2}{*}{Size of DoE}  & & \multicolumn{4}{c}{RRMSE} & & \multirow{2}{*}{\begin{tabular}[c]{@{}c@{}}Computational cost [s]\\ (Excluding simulation cost)\end{tabular}}\\ \cline{4-7}%\cline{3-3}\cline{5-6}\cline{8-9}
 & & & Minimum & Maximum & Mean & Standard deviation & &  \\ \hline
ROM-II  &400 & & 3.2987E-04 & 3.0000E-03 & 8.1723E-04 & 2.6191E-04 & & 2.73 \\
ROM-II & 10000 & & 2.5521E-05 & 5.2281E-04 & 8.6826E-05 & 5.0008E-05 & & 430\\
POD--PCE  & 400 & & 5.9852E-06 & 5.7499E-04 & 7.7981E-05 & 5.8665E-05 & & 5.27\\ \hline
\end{tabular}
\caption{Comparison of computational cost and performance of ROM-II and POD--PCE for the stochastic Ackley function.}\label{Ackley_cost}
\end{table}

From the presented results, one can learn that under the same DoE of size $N_s=400$, the computational costs of the ROM-II for all the examples are much lower than the proposed POD--PCE, however, the accuracy performances are just the opposite. To obtain a comparable accuracy as the POD--PCE method, more simulations of the full order model are needed. For the Ackley function, $9600$ more simulations are needed, which considerably increases the computational cost, even for the ROM-II itself, the computational time excluding the simulation cost is much larger than that of the POD--PCE method. Similar observation can be made for the other two examples. For heat-driven cavity flow and heat-diffusion problem, $600$ and $2400$ more simulations are needed, respectively. Furthermore, one simulation cost of these two problems is much higher compared with the approaches cost. Therefore, the proposed POD--PCE is more efficient compared with the ROM-II method.

\begin{table}[t]
\small
\centering
\begin{tabular}{ccccccccc}
\hline
\multirow{2}{*}{Method} &\multirow{2}{*}{Size of DoE}  & & \multicolumn{4}{c}{RRMSE} & & \multirow{2}{*}{\begin{tabular}[c]{@{}c@{}}Computational cost [s]\\ (Excluding simulation cost)\end{tabular}}\\ \cline{4-7}%\cline{3-3}\cline{5-6}\cline{8-9}
 & & & Minimum & Maximum & Mean & Standard deviation & &  \\ \hline
ROM-II   &400 & & 2.1777E-05 & 1.0900E-02 & 3.4103E-04 & 6.5481E-04 & & 0.573 \\
ROM-II  &1000 & & 1.0370E-05 & 1.0700E-02 & 2.6235E-04 & 5.4645E-04 & & 1.78 \\
POD--PCE & 400 & & 1.6801E-06 & 1.0200E-02 & 2.4889E-04 & 5.2770E-04 & & 4.90\\ \hline
\end{tabular}
\caption{Comparison of computational cost and performance of ROM-II and POD--PCE for the stochastic heat-driven cavity flow.}\label{heat-driven}
\end{table}

%%%%%%%%%%%%%%%%%%%%%%%%%%%%%%%%%%%%%%%%%%%%%%%%%%%%%%%%%%%%%%%%%%%%%%%%%%%
%\subsubsection{Two-dimensional heat diffusion with stochastic conductivity}
\begin{table}[t]
\small
\centering
\begin{tabular}{ccccccccc}
\hline
\multirow{2}{*}{Method} &\multirow{2}{*}{Size of DoE}  & & \multicolumn{4}{c}{RRMSE} & & \multirow{2}{*}{\begin{tabular}[c]{@{}c@{}}Computational cost [s]\\ (Excluding simulation cost)\end{tabular}}\\ \cline{4-7}%\cline{3-3}\cline{5-6}\cline{8-9}
 & & & Minimum & Maximum & Mean & Standard deviation & &  \\ \hline
ROM-II  & 400 & & 0.0000E-00 & 1.3000E-03 & 4.0821E-04 & 2.6473E-04 & & 0.161 \\
ROM-II  & 2800& & 0.0000E-00 & 9.7399E-04 & 1.5943E-04 & 1.8109E-05 & & 11.9 \\
POD--PCE & 400  & & 0.0000E-00 & 9.7555E-04 & 1.5968E-04 & 1.8535E-05 & & 4.93 \\ \hline
%\multirow{2}{*}{Method} & & \multicolumn{4}{c}{RRMSE} & & \multirow{2}{*}{\begin{tabular}[c]{@{}c@{}}Computational cost [s]\\ (Excluding simulation cost)\end{tabular}}&\multirow{2}{*}{Size of DoE} \\ \cline{3-6}%\cline{3-3}\cline{5-6}\cline{8-9}
% & & Minimum & Maximum & Mean & Standard deviation & & & \\ \hline
%ROM-II  & & 0.0000E-00 & 1.3000E-03 & 4.0821E-04 & 2.6473E-04 & & 0.161 & 400 \\
%ROM-II  & & 0.0000E-00 & 9.7399E-04 & 1.5943E-04 & 1.8109E-05 & & 11.9& 2800 \\
%POD--PCE  & & 0.0000E-00 & 9.7555E-04 & 1.5968E-04 & 1.8535E-05 & & 4.93 & 400 \\ \hline
\end{tabular}
\caption{Comparison of computational cost and performance of ROM-II and POD--PCE for the stochastic heat diffusion.}\label{heat-diffusion}
\end{table}

%%%%%%%%%%%%%%%%%%%%%%%%%%%%%%%%%%%%%%%%%%%%%%%%%%%%%%%%%%%%%%%%%%%%%%%%%%%
%\subsubsection{Heat-driven cavity flow with a stochastic boundary temperature}Furthermore, one simulation of heat-driven flow takes about $20$ minutes.
%%%%%%%%%%%%%%%%%%%%%%%%%%%%%%%%%%%%%%%%%%
%%%%%%%%%%%%%%%%%%%%%%%%%%%%%%%%%%%%%%%%%

%%%%%%%%%%%%%%%%%%%%%%%%%%%%%%%%%%%%%%%%%
%\section*{References}

\end{document}